\documentclass[aps, prd, amsmath, floats, floatfix, onecolumn, nofootinbib]{revtex4-2}
\usepackage{graphicx}
\usepackage{latexsym}
\usepackage{color}
\usepackage{dsfont}
\usepackage{amsfonts,amsbsy}
\usepackage[utf8]{inputenc}
\usepackage[mathscr]{euscript}
\usepackage{ amssymb }
\usepackage{amsthm}
\usepackage{amsmath}
\usepackage[english]{babel}
\usepackage[T1]{fontenc}
\usepackage[utf8]{inputenc}
\usepackage{mathtools}
\usepackage{physics}
\usepackage{hyperref}

\newcommand{\be}{\begin{eqnarray}}
\newcommand{\ee}{\end{eqnarray}}
\graphicspath{ Fig 2\ }

\newcommand{\nn}{\nonumber\\}

\theoremstyle{definition}

\def\beq{\begin{eqnarray}}
\def\eeq{\end{eqnarray}}
\def\nn{\nonumber }                                 
\def\r{\ref}                                                

\def\ln{\,\mbox{log}\,}

\def\Tr{\,\mbox{Tr}\,}

\def\al{\alpha}
\def\be{\beta}

\def\ga{\gamma}
\def\de{\delta}
\def\vp{\varepsilon}

\def\ka{\kappa}
\def\la{\lambda}
\def\na{\nabla}
\def\pa{\partial}

\def\si{\sigma}
\def\om{\omega}

\def\th{\theta}

\def\Ga{\Gamma}

\def\La{\Lambda}

\def\B1{ Bianchi-I }
\def\RG{renormalization group }

\def\RN{renormalizable}

\begin{document}

\title{Renormalization Group in Six-derivative Quantum Gravity}

\author{Les\l{}aw Rachwa\l{}}
\email{grzerach@gmail.com}

\affiliation{Departamento de F\'{i}sica---Instituto de Ci\^{e}ncias Exatas,
Universidade Federal de Juiz de Fora, 33036-900, Juiz de Fora, MG, Brazil}

\author{Leonardo Modesto}
\email{lmodesto@sustech.edu.cn}

\affiliation{Department of Physics, Southern University of Science and Technology
(SUSTech), Shenzhen 518055, China}

\author{Aleksandr Pinzul}
\email{aleksandr.pinzul@gmail.com}

\affiliation{Instituto de F\'{i}sica, Universidade de Bras\'{i}lia, 70910-900,
Bras\'{i}lia, DF, Brazil}

\author{Ilya L. Shapiro}
\email{ilyashapiro2003@ufjf.br}

\affiliation{Departamento de F\'{i}sica---ICE,
Universidade Federal de Juiz de Fora, 33036-900, Juiz de Fora, MG, Brazil}

\begin{abstract}

The exact one-loop beta functions for the four-derivative terms (Weyl
tensor squared, Ricci scalar squared and the Gauss-Bonnet) are derived
for the minimal six-derivative quantum gravity (QG) theory in four
spacetime dimensions. The calculation is performed by means of
the Barvinsky and Vilkovisky generalized Schwinger-DeWitt
technique. With this result we gain, for the first time, the full set
of the relevant beta functions in a super-renormalizable model of
QG. The complete set of renormalization group (RG) equations,
including also those for the Newton and the cosmological constant,
is solved explicitly in the general case and for the six-derivative Lee-Wick
(LW) quantum gravity proposed in a previous paper by two of the
authors. In the ultraviolet regime, the minimal theory is shown to be
asymptotically free and describes free gravitons in Minkowski or
(anti-) de Sitter ((A)dS) backgrounds, depending on the initial conditions for the
RG equations. The ghostlike states appear in complex conjugate
pairs at any energy scale consistently with the LW prescription.
However, owing to the running, these ghosts may become
tachyons. We argue that an extension of the theory that involves
operators cubic in Riemann tensor may change the beta functions
and hence be capable of overcoming this problem.
\end{abstract}

\maketitle

\section{Introduction}

Higher derivative terms play an important role in quantum gravity
(QG), as they are required to provide renormalizability (see, e.g.,
\cite{OUP} for introduction). However, including higher derivatives
leads to the presence of ghosts or ghostlike states, representing
a fundamental and, generally, unsolved problem. In the effective
field theory approach, it is usually assumed that the higher derivative
terms in both the classical action
 and quantum corrections are taken
as small perturbations (see, e.g., Ref. \cite{Burgess:2003jk} and
references therein), regardless of the related inconsistencies,
as it was recently discussed, e.g., in \cite{ABSh1}. One of the main problems
 is that the effective approach is applicable only at low
energies, while QG is supposed to cover also the UV region.
An alternative way is trying to provide a consistent treatment of
higher derivative terms in QG at the fundamental level\footnote{One
can also get higher
derivatives within the  spectral action approach to QG \cite{MPR2020}.}.
One of the important aspects, in this case, is the analysis of loop
corrections and of their possible implications on the nature of
physical degrees of freedom of the theory \cite{salstr,Tomboulis-77}.

In the present paper, we consider the higher derivative operators in
the classical action as a viable {\em nonperturbative} completion
of Einstein's gravity to achieve a super-renormalizable theory.
In other words, the higher derivative operators here are not small in
comparison to the Einstein-Hilbert action, but actually dominant in
the ultraviolet regime, contrary to the effective approach. A usual
proposal to overcome the nonrenormalizability of Einstein's theory
is to introduce higher derivative terms, with four or more derivatives
of the metric, improving the UV convergence of the theory at
quantum level in the perturbative loop expansion.

The first proposal was the strictly renormalizable quadratic in curvature
theory of gravity \cite{Stelle}, while super-renormalizable and finite
theories for quantum gravity were proposed later on in
\cite{krasnikov,kuzmin,highderi,Tomboulis-97}
and  \cite{Modesto:2011kw,Modesto:2014lga,Modesto:2017sdr,LM-Sh,
ModestoLeeWick}. As we already
mentioned above, the price to pay is the introduction of extra degrees of
freedom in the spectrum of a renormalizable theory, including real and/or
complex ghostlike particles. Recently, it has been shown that higher
derivative theories can be formulated as perturbatively unitary
\cite{AnselmiPiva1,AnselmiPiva2,AnselmiPiva3}, if implemented via
a completion of the prescription introduced by Cutkosky, Landshoff,
Olive, and Polkinghorne \cite{CLOP} for the Lee-Wick theories
\cite{LW1, LW2}. In this case, the ghost fields are harmless for
unitarity and can only exist as virtual particles, namely they are
never onshell.
One can also compare the situation in higher derivative models
with Ho\v{r}ava-Lifshitz models, where one also has up to six derivatives
on the metric tensor, but they are only spatial derivatives (see, e.g.,
\cite{Pinzul:2010ct}). Therefore, one can view that the price to pay is
either to have perturbative ghosts or to end up with Lorentz symmetry
violation. Here we focus on the first possibility.

In this paper, we will consider the minimal six-derivative gravity
theory and we will compute the divergent contributions to the
quantum effective action. Afterwards, given all the beta functions
of the theory, we will study the renormalization group flow to infer
about the ultraviolet behavior of gravity.

The derivation of one-loop divergences and renormalization group
equations has been for long a needful part of the study of models
of quantum gravity (QG). The first such calculations were done for
the QG based on Einsteinian general relativity with matter in the seminal papers
\cite{hove,dene}. This calculation confirmed that the theory is not
renormalizable, but the one-loop $S$-matrix in the pure QG is finite,
since the one-loop divergences vanish on shell\footnote{Notice that in
the case of QG interacting with matter fields or any kind of external
$T_{\mu\nu}$  the finiteness of the $S$-matrix does not hold.}.  The
analysis of the dependence on the choice of the gauge-fixing started in
\cite{KTT} and was developed further including the nontrivial
parametrizations of the quantum field \cite{Kalm,FirPei} and quite
recently in \cite{JQG} (see also  \cite{RP-2017}) and the first
order (Palatini) formalism \cite{BuSh1,BuSh2}. The two-loop
calculations were done in \cite{gorsag} and \cite{vandeVen} and
confirmed that there is no chance to have a renormalizable QG based
on Einstein's gravity.

The one-loop calculations in the four-derivative gravity
\cite{Stelle} have been done in \cite{JuTo,frts82,avbar86} for the
general and in \cite{frts82,AMM-92} for the conformal versions of
the theory, with an effective verification of calculations in
\cite{Gauss}. These one-loop calculations confirmed the
expectations to meet a renormalizable QG in this case. The
gauge-fixing dependence in this case \cite{frts82,a} is not so strong
as in the quantum Einstein's theory of gravity, but it is still impossible
to arrive at the well defined renormalization group equations for the
cosmological constant $\La$ and the Newton constant $G$\footnote{The remarkable exception is the effective
framework based on the Vilkovisky-DeWitt unique effective
action \cite{TV90,UEA-QG}.}. The renormalization of all higher
derivative terms (Weyl squared $C^2$, Ricci scalar squared $R^2$
and the Gauss-Bonnet term $E_4$) does not have this problem.
The \RG equations for these parameters have
interesting applications, e.g. the cosmological and astrophysical
models with running $\La$ \cite{CCnova} or $G$ \cite{RotCurves}.
Finally, the \RG for the $R^2$ term can be relevant for establishing
the particle physics background in the Starobinsky model of
inflation \cite{star}, as has been recently discussed in \cite{StabInstab}.

The set of universal, gauge-fixing- and parametrization-independent \RG
equations for all effective parameters can be obtained within the
super-renormalizable models of QG, that require more than four
derivatives. The early suggestions to construct super-renormalizable
QG \cite{krasnikov,kuzmin} (see the latest developments in
\cite{Modesto:2011kw, Modesto:2014lga, Modesto:2017sdr})
were based on nonlocal actions with an infinite number of derivatives.
The evaluation of power counting in this case is well defined only for
asymptotically polynomial nonlocal form factors and the
super-renormalizable nature of these models can be confirmed
\cite{CountGhost}. The problem of dependence on the gauge-fixing
 parameters can be easily solved  for the super-renormalizable
QG theories based on polynomial actions with six or more derivatives.
Indeed, for local as well as analytic nonlocal theories, one can
prove that the \RG equations for all parameters do not depend on
the choice of the gauge-fixing \cite{highderi}.

Owing to the power counting super-renormalizability, the nonlocal
theories of QG have universal, i.e., gauge- and
parametrization-independent beta functions. On top of that, these
theories can be formulated as unitary
\cite{Briscese:2018oyx} and the stability
\cite{Briscese:2018bny,Briscese:2019rii} is secured at any
perturbative order. At the quantum level, the feature of reflection
positivity has been discussed in the different frameworks in
\cite{ABSh1} and \cite{Christodoulou:2018jbn}.

Here we emphasize that the stability is a more general concept than perturbative unitarity. The
latter can be precisely and rigorously formulated only around flat spacetime and for scattering
amplitudes, so for the elements of $S$-matrix. Around a general nonflat background (precisely such
background which is not asymptotically flat) in gravitational theory we cannot uniquely formulate the
scattering problem, define asymptotic free and noninteracting states and discuss unitarity of the
scattering matrix. Still on a general background we can discuss the stability properties of the theory.
As it is usually done in gravitational contexts here we mean the stability of small perturbations of
equations of motion (EOM)
analyzed around a given classical background, so here we quadratize the action and for example look
for spacetime evolution of perturbations and can find Lyapunov exponents of the stability (both for the
perturbative and nonperturbative levels of changes of the background metric
\cite{Briscese:2018bny,Briscese:2019rii,Christodoulou:2018jbn}). This notion of stability is, of course,
different than the issue of stability of ground states of the quantum system describing vacua of these
quantum field theory (QFT) models. As it is well known the latter concept is investigated by truly nonperturbative methods, for
example, of vacua tunneling and spontaneous symmetry breaking in the ground states of these QFTs.

The unitarity and stability issues have been considered also in the
polynomial higher derivative QG. In particular, in the recent papers
\cite{LM-Sh}
and \cite{ModestoLeeWick}, it was shown that the QG model with six
derivatives is tree-level unitary if the massive poles of the classical
propagator form complex conjugate pairs. The power counting in
these models guarantees super-renormalizability and, therefore, the
polynomial QG models also have universal running of all parameters,
in the same sense as the nonlocal models have.

The difference between nonlocal and polynomial theories of QG is
that, in the nonlocal case, there is no available reliable covariant technique
for the one-loop calculations. On the other hand, in the polynomial
models, the one-loop divergences can be obtained using Barvinsky and
Vilkovisky universal traces \cite{bavi85}, albeit such calculations
may be rather cumbersome and technically nontrivial. Until now,
the list of the known beta functions included the ones for $\La$
and $G$, derived in \cite{highderi} and \cite{SRQG-beta},
respectively.
In the present work, we report on the quantum calculations of the
remaining beta functions in a minimal six-derivative model of QG,
which includes the versions with real and  complex poles
 corresponding to both normal particles and ghosts.
 The last case is the one which was used in \cite{LM-Sh} for the
Lee-Wick analysis of the unitarity. Indeed, the beta functions
obtained in what follows form, together with the previous ones
\cite{SRQG-beta}, the full set and, therefore, enable one to
explore how the running affects the positions of the poles of
the propagator of metric perturbations on the complex plane.
From this perspective, the derivation of the full set of equations
describing the running is the necessary step in describing the
theory at the quantum level.

From a technical point of view, the one-loop calculations in
super-renormalizable QG present more difficulties compared to
the ones in the four-derivative renormalizable models
\cite{JuTo,frts82,avbar86}. The level of complexity of such
a calculation depends on the type of one-loop counterterms. The
counterterm for the cosmological constant is actually easy to
obtain \cite{highderi}. However, the derivation of the divergence
linear in the scalar curvature requires really big efforts and was
done only recently in \cite{SRQG-beta}.
In the present work, we make the next step, calculating the simple
one-loop divergences for the four-derivative sector. As a result,
we arrive at the beta functions for the Weyl-squared,  $R^2$
and Gauss-Bonnet terms. The calculation is really cumbersome
and it is done for the simplest possible six-derivative theory
without ``killer'' terms in the classical action, which here are third
powers of the generalized curvature tensor $\cal R$. Even in this simplest case,
the intermediate expressions are too large for the presentation,
hence they will be mostly omitted. Similar computations in four-, six- and general
higher derivative gauge theory were performed in \cite{Babelon:1980bu,Asorey:1995tq,Modesto:2015lna,Modesto:2015foa,Asorey:2020omv}.

The work is organized as follows.
In Sec.~\ref{s2} we briefly review the renormalization in the six-derivative
 model of QG and show that the divergences in this
theory do not depend on the gauge-fixing. Section ~\ref{s3} is devoted
to the derivation of one-loop divergences, while in Sec.~\ref{s4} we
introduce the renormalized Lagrangian, the beta functions for the
four-derivative operators and the related \RG equations. In
Sec.~\ref{s5} we apply our results to the six-derivative Lee-Wick
quantum gravity solving exactly the \RG equations for all running
coupling parameters. Later in Sec.~\ref{s6} we discuss the
issue of asymptotic freedom in this model.
Finally, in Sec.~\ref{s7} we draw our conclusions and give a little discussion of
additional related issues.
In the special appendix \hyperref[app]{A} we  present a detailed analysis of the Wick rotation
in local higher derivative quantum field theories.

\section{Power counting and gauge-fixing-independence of divergences}
\label{s2}

The action of the theory that we consider in this paper has the form
\beq
S_{\rm QG} =
\!\int\! d^4x \sqrt{|g|} \,{\cal L} \, ,
\label{6model}
\eeq
where the density of Lagrange function
reads\footnote{The list of all covariant six-derivative terms
can be found in Ref.~\cite{Bonora}, where it was developed as a
part of the study of conformal anomaly in six spacetime dimensions.}
\beq
{\cal L} =  \om_C C_{\mu\nu\!\rho\sigma} \square C^{\mu\nu\!\rho\sigma}
+ \om_R R \square R
+ \th_C  C^2
+ \th_R R^2
+  \th_{\rm GB} E_4
+ \omega_\kappa R + \omega_\Lambda
\, .
\label{L}
\eeq
Here $C_{\mu\nu\!\rho\sigma}$ is the Weyl tensor and $E_4$ is the
Gauss-Bonnet term, which is the Euler characteristic in $d=4$,
\beq
&&
C^2 \,=\,C_{\mu\nu\!\rho\sigma}^2
\,=\, R_{\mu\nu\!\rho\sigma}^2 - 2 R_{\mu\nu}^2 + \frac13\,R^2,
\\
&&
E_4 \,=\,
R_{\mu\nu\!\rho\sigma}^2 - 4 R_{\mu\nu}^2 + R^2.
\label{WGB}
\eeq
In what follows we will use the pseudo-Euclidean
notations and $\sqrt{|g|}$ denotes the absolute value of the square root
of the metric determinant. The coupling $\omega_\kappa$ is related to the Newton constant $G$, while
$\omega_\Lambda$ to the cosmological constant $\Lambda$.
The theory with the Lagrangian (\r{L}) is the simplest one that
describes the general form of the graviton propagator around flat
 spacetime in four spacetime dimensions with
six derivatives. It is also the simplest QG model  admitting, besides
the massless graviton, only Lee-Wick complex conjugate poles
at the classical level \cite{LM-Sh}.

Let us note, from the very beginning, that there are two remarkable
special cases in the theory (\ref{L}).
In order to have a nondegenerate classical
action one needs that both coefficients $\om_C$ and $\om_R$
should be nonzero and the quantum calculations reported in the
next section correspond only to this kind of model.
Correspondingly,
in the special case of  $\om_C=0$ and $\om_R\neq 0$,
the theory has the propagating spin-two mode with four derivatives
and the propagating spin-zero mode with six derivatives, while
interaction vertices have six derivatives in both special and generic
theories. For another special version, with $\om_C\neq 0$ and
$\om_R= 0$, the situation is opposite, but according to the
power counting arguments
\cite{highderi} in both special cases the theory is nonrenormalizable.

In the case of both $\om_C$ and $\om_R$ nonzero, the theory
(\r{L}) is the simplest example of a super-renormalizable model of
QG. The power counting procedure results \cite{highderi} in
the expression:
\beq
D+d = 6-2p,
\label{1}
\eeq
where $p$ is the number of loops of the diagram,
$d$ is the number of derivatives of the metric on the external lines
or, equivalently, the mass dimension of the divergences, $D$ is the
superficial degree of divergences of the diagram. For the logarithmic
divergences, $D=0$ and we can see that the one-loop divergences
have $d=4$, while higher loops give less derivatives in the
counterterms. This result has deep consequences for the general
structure of renormalization in the model (\r{L}). Let us list
here these aspects of the theory.
\vskip 1mm

{i}. \ There are no UV divergences with $d=4$ at higher loop orders
 than the first one. Thus, the  beta functions for the four-derivative
  terms, which can be derived
 at one-loop order, are exact and hold to all perturbative orders in
 the loop expansion.  The situation is different for the Einstein
 term with $d=2$, which gains divergent contributions  (and
 hence contributions to the beta functions) for $p=2$ and for the
 cosmological constant
 term which gains such contributions for both  $p=2$ and  $p=3$.
 Starting  from $p=4$ (four-loop order) the loop corrections are
 finite, i.e., the theory is super-renormalizable.
\vskip 1mm

{ii}. \  The terms with four, two and zero derivatives in the action
(\r{6model}) do not affect the counterterms with $d=4$. This means
that the practical calculations of the $d=4$ counterterms in this
theory can be performed for the reduced model with the action
\beq
S_{{\rm 6der}} =
\!\int\! d^{4}x\sqrt{|g|}
\big\{
 \om_{C}\,C_{\mu\nu\!\rho\sigma}\square C^{\mu\nu\!\rho\sigma}+\,\,
\om_{R}\,R\square R\big\}.
\label{6red}
\eeq
On the other hand, the  $d=4$ counterterms may be affected
by inclusion of extra $O\left({\cal R}^3\right)$-type terms
into the action. These terms are not necessary, e.g., for
renormalizability and should be regarded as nonminimal.
The practical calculations in the next section were performed
without these terms.
\vskip 1mm

{iii}. \  The power counting (\r{1}) shows
that the classical action and the divergences have different
numbers of derivatives of the metric. This feature implies that
the theory is super-renormalizable, but also results in the
universality of quantum corrections that we mentioned in
the Introduction. According to the general statements about
gauge-fixing and parametrization dependence of the effective action (see,
e.g., \cite{VLT82} for a general proof and \cite{book} for the
reduced simplified version valid for the one-loop approximation),
both kinds of dependencies disappear on the mass shell. If we
describe the ambiguity in gauge and parametrization by the
parameters $\al_i$ and use the logic of \cite{frts82} (and, more
explicitly, in \cite{a,JQG}),  then the difference between the
effective actions $\Gamma$ for the two choices of these
parameters, namely $\al_i$ and $\al_i^{0}$, can be written
as follows:
\begin{equation}
\Ga\big(\al_i\big) \,-\, \Ga\big(\al^{0}_i\big)
\,=\,
\!\int\! d^4x\sqrt{|g|}\,\,\vp^{\mu\nu}\,f_{\mu\nu},
\label{2}
\end{equation}
where $\,f_{\mu\nu}=f_{\mu\nu}(g_{\kappa\lambda},\al_i,\al_i^{0})\,$
is an unknown function of the metric and parameters $\al_i$
and $\al_i^0$ and
\begin{equation}
\vp^{\mu\nu}
\,=\,\frac{1}{\sqrt{|g|}} \,\frac{\de \Ga(\al_i^{0})}{\de g_{\mu\nu}}.
\label{3}
\end{equation}
If we are interested in the divergent part of the effective action,
$\,f_{\mu\nu}\,$ is a local expression which is at least of the first
order in the loop expansion parameter $\hbar$. Thus the dimension
of the right-hand side of (\r{2}) is at least equal to the
dimension of $\,\vp^{\mu\nu}$. Furthermore, the equations of motion
$\,\vp^{\mu\nu}\,$ can be expanded into a power series in $\hbar$.
At the lowest level of this expansion, one meets classical
equations of motion, hence $\,\vp^{\mu\nu}$ is at least of the
sixth order in the metric derivatives. At the same time,
according to the power counting (\r{1}), the divergent part of the
left-hand side of (\r{2}) contains only four derivatives of the metric.
Therefore, the divergent parts of the both sides of Eq. (\r{2})
vanish implying that $f_{\mu\nu}^{\rm div}=0$.
The conclusion is that the right-hand side of (\r{2}) is a finite
expression and, hence, the divergent part of the effective
action $\Ga$ does not depend on the gauge-fixing or
parametrization ambiguities.

All in all, we have shown that the beta functions of the theory
(\ref{L}) are defined in a universal way, that means they
are free of ambiguities which are typical for QG based on
Einsteinian general relativity or on the \RN \ four-derivative theory of
gravity. As it was already mentioned above, the derivation of zero-
and two-derivative divergences has been previously done in Refs.
\cite{highderi} and \cite{SRQG-beta}. In what follows we will
complete the set of beta functions for the theory (\ref{L}) by
deriving the exact beta functions for the four-derivative
coefficients $\,\theta_C$, $\,\th_R$ and $\, \th_{\rm GB}\,$. Without
loss of generality the calculation will be performed for the reduced
model (\r{6red}).

\section{One-loop calculation in six-derivative model}
\label{s3}

Except for a larger volume of algebra, the one-loop calculations
are performed using the same general algorithm developed in the
super-renormalizable models  \cite{SRQG-beta}. Thus we can
skip a great part of the explanations and give only specific details
of the calculation of the UV-divergent four-derivative terms.

The background field method is defined by splitting the metric
\beq
g_{\mu\nu} & \longrightarrow & g_{\mu\nu} + h_{\mu\nu}.
\label{backgr}
\eeq
Since the result does not depend on the gauge-fixing, the parameter $\be$ in a linear gauge-fixing condition $\chi_\mu$,
\beq
\chi_{\mu} =
\na^\la h_{\la\mu} - \be \, \na_\mu h, \qquad
h =  h^{\nu}{}_{\nu} \, ,
\label{chi}
\eeq
can be chosen in the most simple ``minimal'' way, as we will show
below. The same concerns the parameters $\al$ and $\ga$ in the
weight operator $\,\hat{C}=C^{\mu\nu}$,
\beq
C^{\mu\nu}\,=\,-\frac{1}{\al}\left(
g^{\mu\nu}\square^{2} + (\ga-1)\na^{\mu}\square\na^\nu\right).
\label{weight}
\eeq
This operator is already in a self-adjoint form and it is originally defined when acting on a covariant vector $\chi_\nu$.
Together with the gauge-fixing $\chi_\mu$
the operator $\hat{C}$ defines the gauge-fixing action \cite{frts82},
\beq
S_{{\rm gf}}=\!\int\! d^{4}x\sqrt{|g|}\,\,\chi_\mu \,C^{\mu\nu}\,
\chi_\nu .
\label{gf}
\eeq
The action of the Faddeev-Popov complex ghosts has the form
\beq
S_{{\rm gh}}
\,=\,
\!\int\! d^{4}x\sqrt{|g|}\,\, \bar{C}^{\mu}M_{\mu}{}^{\nu}C_{\nu},
\label{gh}
\eeq
where the bilinear part depends on $\chi_\mu$ and on the
generator of gauge transformations,
\beq
{\hat  M}\,=\,
M_{\mu}{}^{\nu}
\,=\,\frac{\de \chi_\mu}{\de g_{\al\be}}\,R_{\al\be}\,^\nu
\,=\,\de_\mu^\nu \square
+ \na^\nu \na_\mu - 2\be\na_\mu \na^\nu.
\label{M}
\eeq

Let us give a few more details concerning the choice of the gauge-fixing
parameters.  The bilinear form of the action is defined from the
second variational derivative,
\beq
{\hat H}
\,=\,
H^{\mu\nu,\rho\si}
\,=\,
\frac{\de^2 \left(S_{\rm QG} + S_{{\rm gf}}
\right)}{\delta h_{\mu\nu}\,\delta h_{\rho\sigma}}
\,=\,H_{{\rm lead}}^{\mu\nu,\rho\si} \,+\,
O({\cal R}),
\label{bil}
\eeq
where the first term includes six-derivative terms and
$\,{O}({\cal R})\,$ is the rest of the bilinear form with
four of less derivatives and nonvanishing powers of curvature $\cal R$. The
dimension is compensated by the powers of curvature tensor and its
covariant derivatives; we can also commute
some covariant derivatives and trade them to curvature
 in this
case\footnote{One can consult Appendix A of \cite{MPR2020} for
explicit forms of such relations.}. The
corresponding terms are very bulky in the present case and it is
not reasonable to include them in the paper.

The six-derivative part of the ${\hat H}$ operator, after adding the
gauge-fixing term, has the form
\beq
H_{{\rm lead}}^{\mu\nu,\rho\si}
&=&
\Big[ \frac{\om_C}{2}\,\left(g^{\mu\rho}g^{\nu\si}+g^{\mu\si}g^{\nu\rho}\right)
\,+\, \Big(
- \frac{\om_C}{3}
+ 2\om_{R}+ \frac{\be^{2}\ga}{\al}\Big) g^{\mu\nu} g^{\rho\si}\Big] \square^3
\nn
\\
&&
+\,\,
\Big( \frac{\om_C}{3} - 2\om_R - \frac{\be\ga}{\al}\Big)
\Big( g^{\rho\si}\na^{\mu}\na^{\nu}
+ g^{\mu\nu}\na^{\rho}\na^{\si} \Big) \square^{2}
\nn
\\
&&
+\,\,\Big( - 2\om_C +\frac{1}{\al} \Big)g^{\mu\rho}
\na^{\nu}\na^{\si}\square^{2}
+ \Big( \frac{2\om_C}{3} + 2\om_{R} + \frac{\ga-1}{\al}\Big)
\na^{\mu}\na^{\nu}\na^{\rho}\na^{\si}\square \,.
\label{bilin}
\eeq
For the sake of brevity, in this expression we left implicit
symmetrization inside the pairs of indices $(\mu,\nu)$ and $(\rho,\si)$.

It is easy to see that to make the operator ${\hat H}$ minimal, one
has to choose the following values of the gauge-fixing parameters
\cite{SRQG-beta}:
\beq
\alpha=\frac{1}{2\omega_{C}},
\qquad
\beta=\frac{\om_{C}-6\om_{R}}{4\om_{C}-6\om_{R}},
\qquad
\gamma=\frac{2\omega_{C}-3\omega_{R}}{3\omega_{C}}.
\label{mingauge}
\eeq
As we have explained in the previous section, this choice does
not affect the one-loop divergences in super-renormalizable QG.
Thus, we assume (\ref{mingauge}) as the most simple option.

With these values of the gauge-fixing, Eq.~(\r{bilin}) becomes
\beq
H_{{\rm lead}}^{\mu\nu,\rho\si}
&=&
\om_C
\Big (\de^{\mu\nu , \rho\si} + k g^{\mu\nu}g^{\rho\si}\Big)\square^3 \quad {\rm with }\quad \de^{\mu\nu , \rho\si}=\frac12\left(g^{\mu\rho}g^{\nu\si}+g^{\mu\si}g^{\nu\rho}\right),
\label{Hmin}
\eeq
where the coefficient  $k$  is
\beq
k &=& -\, \frac{\om_{C}-6\om_{R}}{4\om_{C}-6\om_{R}}.
\label{kexp}
\eeq

In the expressions for $\beta$ in (\ref{mingauge}) and for $k$ in
(\ref{kexp}), the denominator blows up near the point where
$\omega_C = \frac{3}{2} \omega_R$. This point corresponds
to the leading six-derivative term of the gravitational action in
(\ref{L}) of the form $R_{\mu\nu}\square R^{\mu\nu}$. However,
the final results show no singular dependence on the
$4\om_{C}-6\om_{R}$. Therefore, the results in this special case
can be regarded as a smooth analytic continuation from the generic
case when $4\om_{C}-6\om_{R}\neq0$.

The DeWitt metric in the internal space of metric fluctuations
$h_{\mu\nu}$ has the form
\beq
{\cal G}^{\mu\nu,\rho\si}
\,=\,
x_1\, \de^{\mu\nu , \rho\si} + x_2\, g^{\mu\nu}g^{\rho\si},
\eeq
where
\beq
x_{1}=\om_{C}
\qquad
\mbox{and}
\qquad x_2 = k \, \om_{C}.
\eeq
\\
The inverse DeWitt metric can be found in the form
\beq
{\cal G}_{\ka\la,\mu\nu}^{-1}
= \frac{y_1}{2}\, g_{\ka\mu }g_{\la\nu}+\frac{y_1}{2}\, g_{\ka\nu }g_{\la\mu} + y_2\, g_{\ka\la}g_{\mu\nu} =
y_1 \,\de_{\ka\la,\mu\nu } + y_2 \,g_{\ka\la}g_{\mu\nu},
\eeq
such that
\beq
{\cal G}_{\ka\la,\mu\nu}^{-1}\,{\cal G}^{\mu\nu,\rho\si} &=&
\de_{\ka\la}\,^{\rho\si}
\,=\,
\frac12
\Big(
\de_{\ka}^{\rho}\,\de_{\la}^{\si}\,+\,\de_{\ka}^{\si}\,\de_{\la}^{\rho}
\Big).
\eeq
The coefficients can be easily found to be
\beq
y_{1} = \frac{1}{\om_{C}},
\qquad
y_{2} = \frac{\om_{C}-6\om_{R}}{18\om_{C}\om_{R}}\, .
\eeq
In the expression for $y_2$ we explicitly see that limits $\om_C\to0$ and $\om_R\to0$ are singular.

We find that new operator $H'$ is in the standard minimal form
\beq
{\hat H}'
&=&
H'_{\ka\la}{}^{\rho\si}
\,=\,
{\cal G}_{\ka\la,\mu\nu}^{-1}\,H^{\mu\nu,\rho\si}
\,=\,\de_{\ka\la}{}^{\rho\si}\square^{3}\,+\,O({\cal R}) \, .
\eeq
For the divergent part we have $\,\Tr \ln {\hat H}' = \Tr \ln {\hat H}$.

Now we have all necessary elements to write down the general
formula for the one-loop contribution to the effective action of
the theory,
\beq
{\Ga}^{(1)}
\,=\,\frac{i}{2} \Tr \ln {\hat H}
\,-\,i \Tr\ln  {\hat  M}
\,-\, \frac{i}{2}\Tr \ln  {\hat C}.
\label{TrLn}
\eeq

The calculation of the divergent parts of the first two expressions
in (\r{TrLn}) is pretty much standard \cite{SRQG-beta}, with the use of the generalized
Schwinger-DeWitt technique \cite{bavi85}. For this reason we
shall skip most of the technical details and will just comment
on the derivation of the last term.

In order to derive the $O\left({\cal R}^2\right)$ terms in the
divergent part of effective action (\ref{TrLn}) coming from the operator $\hat C$ in (\ref{weight}),
we remind the reader that
both degenerate $\,\ga=0\,$ and nondegenerate $\ga \neq 0$ versions of it
are possible, according to Eq.~(\ref{mingauge}). These two cases
should be considered separately. For the sake of simplicity, we focus
only on the nondegenerate case. It turns out to be useful to introduce
another conjugate operator $\hat B$, whose contribution can be found in \cite{bavi85},
\beq
B_{\nu\la}\,=\,-\al\left(g_{\nu\la}\square + b\na_\nu \na_\la\right)\,.
\label{B}
\eeq
One can choose the parameter $b$ in such a way that the product of
two operators is a minimal sixth-order operator,
\beq
H^{\mu}{}_{\lambda}
= C^{\mu\nu}B_{\nu\lambda} =
\delta^{\mu}_{\lambda}\square^{3} +  O\left(\cal R\right).
\label{H20}
\eeq
The value of the parameter $b$ is obtained from the last condition
(\ref{H20}), namely
\beq
C^{\mu\nu}B_{\nu\lambda}
&=&
\delta^{\mu}_{\lambda}\square^{3}
+ b\nabla^{\mu}\nabla_{\lambda}\square^{2}
+ (\ga-1)\nabla^{\mu}\nabla_{\lambda}\square^{2}
+ b(\ga-1)\nabla^{\mu}\nabla_{\lambda}\square^{2}
+ O\left(\cal R\right).
\eeq
To cancel the nonminimal terms we need
\beq
b = \frac{1- \ga}{\ga}.
\eeq
Starting from (\ref{H20}), we find
\beq
&&
\Tr \ln  \left( C^{\mu\nu}B_{\nu\lambda} \right)
\,=\,
\Tr \ln C^{\mu\nu} + \Tr \ln B_{\nu\la}
\nn
\\
&&
\quad
=\,
\Tr \ln
\big\{
\de^\mu_\la \square^3
+ (V^\mu{}_\la)^{\rho\si\tau\om}\na_\rho\na_\si\na_\tau\na_\om
+ (U^\mu{}_\la)^{\rho\si}\na_\rho\na_\si  + \ldots  \big\} \,,
\label{BCprod}
\eeq
where $U$ and $V$ tensors are with nonzero powers of curvature. The last expression, in the right-hand side, can be elaborated using the universal
traces in \cite{bavi85},
while the formula for the divergent part of $ \Tr \ln B_{\nu\la}$
can be found in the same paper. The ellipsis in (\ref{BCprod}) denotes terms
cubic in curvature or with background dimensions 6, hence not contributing to divergences.

As we have already mentioned above, the intermediate expressions for
our calculations are too cumbersome for the LaTeX file, so they will not
be presented here.

Let us note that the computation of ${\rm Tr}\ln \hat C$ was checked
using three methods. Since the $\hat C$ operator is a nonminimal
four-derivative differential operator with vector indices, then the
computation of its trace of the logarithm is a bit tricky and one
has to be careful. This is why we performed here additional
verifications of our partial results for this trace. All the methods
consist basically of creating new operators, with higher number
of derivatives. By selecting adjustable parameters, those
could be put into a minimal form. This reduction was achieved by
an operatorial multiplication by some two-derivative spin-one
operator $\hat B$ containing one free parameter, e.g., $b$ as in (\ref{BCprod}).

In the first method, we multiplied $\hat C$ by another operator from the
right side. In the second case, the multiplication was from the left.
And, in the third method, we used the symmetric form of multiplication
$\hat B \hat C \hat B$, which is important for the self-adjointness property of the
resulting operator. In this case, $\hat B$ has the form of a two-derivative
operator, whose trace of the logarithm is known (and was also
checked below). We remark that in the first two methods the resulting
operators were six-derivative ones, while in the last one it was an
eight-derivative operator. All three methods of computation of
${\rm Tr}\ln \hat C$ agree for the divergent terms quadratic in curvatures,
appearing in the form of UV divergences of the theory
($R^2$, $R_{\mu\nu}^2$ and $R_{\mu\nu\!\rho\sigma}^2$).

 Similarly, the computation of ${\rm Tr}\ln \hat B$ was checked using three analogous methods.  We used multiplication from  both sides by the operator $\hat A$
and also the symmetric form  $\hat A \hat B \hat A$,
where $\hat B$ is a two-derivative operator, whose trace of the logarithm is searched
for. Here, $\hat A$ is a two-derivative nonminimal spin-one vector gauge operator, whose
trace of the logarithm is well known \cite{frts82,bavi85}. Again, all three methods agree for terms quadratic
in curvatures.

\subsection{The results for the traces and effective action}

Skipping the intermediate details, we present the results for
the functional traces. It proves useful to use the covariant cutoff
regulator $L$ \cite{bavi85}, related to the dimensional regularization parameter $\epsilon$
by \cite{bro-cass,bavi85}
\beq
\ln L^2 \equiv \ln \frac{ \Lambda^2 }{\mu^2}
 = \frac{1}{\epsilon}
=  \frac{2}{4-n},
\label{dimreg}
\eeq
where $n$ is the regularized dimension of spacetime.
The contribution of the main operator $\hat{H}$ to the
divergent part of the expression of our interest (\ref{TrLn}) has the form
\beq
{\rm Tr}\, \ln {\hat H}\Big|_{\rm div}
& = &
\frac{i\ln L^{2}}{(4\pi)^{2}}\!\int\! d^4x \sqrt{|g|}
\bigg\{
\Big( \frac{2x}{9} + \frac{143}{8} \Big)
R_{\mu\nu\!\rho\si}^2
\nn
\\
&-&
\frac{1}{2x-3} \Big( \frac{8x^2}{9}  + \frac{22039\,x}{216} -\frac{943}{6}
+ \frac{33}{8x} - \frac{1}{x^2} \Big)
R_{\mu\nu}^2
\nn
\\
&+&
\frac{1}{2x-3}
\Big(\frac{4x^2}{27} + \frac{17995\,x}{864}
- \frac{1493}{48} - \frac{31}{32\,x} + \frac{1}{2x^2}\Big)
R^2
\Big\},
\label{TrLnH}
\eeq
where we introduced the useful notation
\beq
x=\frac{\om_C}{\om_R}.
\label{x}
\eeq
Some observation is in order at this stage.
From the quantum field theory arguments discussed above we cannot expect that the limit $\,\om_C \to 0\,$
($x \to 0$) will be continuous because when $\,\om_C = 0\,$ exactly, the number of degrees of freedom of the theory changes. But still for every $\de >0$ one might expect that the divergent part of the effective action is continuous and bounded for $x > \de$. Therefore, all the terms proportional to $1/x$ in (\ref{TrLnH}) should cancel out in the final expression. This can be seen in (\ref{Ga}) below as an effective test of the correctness of calculations\footnote{The same intuitive argument cannot be immediately adopted for the other case when $\omega_R\to0$ ($x\to\infty$) as in this case there is a more radical change of the model.}.

The contribution of the bilinear ghost sector is related
to the expression
\beq
{\rm Tr}\,\ln \hat M\Big|_{\rm div}
\,=\,
\frac{i\ln L^{2}}{(4\pi)^{2}}\!\int\! d^4x \sqrt{|g|}\Big\{
\frac{11}{180}\,R_{\mu\nu\!\rho\si}^2
\,-\, \frac{74x^2 - 150 x + 45}{270\,x^2}R_{\mu\nu}^2
-
\frac{19 x^2 - 12x + 9}{108\,x^2} R^2
\Big\}.
\label{TrLnM}
\eeq
And finally, the contribution of the weight operator $\hat C$ from (\ref{weight}) has the form
\beq
{\rm Tr}\ln \hat C\Big|_{\rm div}
\,=\,
\frac{i\ln L^{2}}{(4\pi)^{2}}\!\int\! d^4x \sqrt{|g|}
\Big\{
\frac{11}{90}\,R_{\mu\nu\!\rho\si}^2
- \frac{ 111x^2-154x  + 15}{120x (2x-3)}R^2_{\mu\nu}
+
\frac{11x^2 -14x + 3}{96x (2x-3)}\,R^{2}\Big\}.
\label{TrLnC}
\eeq
Summing up the expressions (\r{TrLnH}), (\r{TrLnM}) and (\r{TrLnC})
according to (\r{TrLn}), we arrive at the final result for the divergent
contribution to the quantum effective action for the operators $C^2$,
$R^2$ and $E_4$,
\beq
\Ga^{(1) C, R, E}_{\rm div}
&=&
-\frac{\ln L^{2}}{2(4\pi)^{2}}\!\int\! d^4x \sqrt{|g|}
\left\{
\Big(\frac{2x}{9}+\frac{397}{40}\Big)C^2
- \frac{7}{36}R^{2}
+ \frac{1387}{180}E_4
 \right\},
\qquad
\label{Ga}
\eeq
where we used a more physical basis defined by Eq.~(\r{WGB}).
It is remarkable that in this expression there are no singular terms
of the type $1/x$, as it was expected.
Note that these terms
are present in the intermediate expressions for (\r{TrLnH}),
(\r{TrLnM}) and (\r{TrLnC}), hence we conclude that their
cancellation in the sum is a partial verification of the result.

On the other hand, the smooth limit $x\to 0$ does not mean that
the result (\ref{Ga})
remains valid for $x=0$. As we explained above, in this case the
form of the operator $\hat{H}$ changes dramatically and the
form of the divergences is supposed to change too. The calculation
in this case can be done, e.g., using the method described in
Chapter $8$ of \cite{book}. At the same time, the theory with
$\om_C=0$ is nonrenormalizable and, hence, such calculation
does not look sufficiently interesting.

The expression (\r{Ga}) confirms the expectations based on power
counting, general covariance and locality. The result is covariant,
local and has four derivatives of the metric,
in agreement with Eq.~(\r{1}). Furthermore, all the
coefficients of the three divergent terms of (\r{Ga}) depend only
on the dimensionless ratio (\r{x}) of the terms with six derivatives
of the classical action, but not on the lower derivative terms, as it
was in the case of the cosmological constant and the linear in $R$
divergences \cite{SRQG-beta}.

Let us say that regardless of the simplicity of the formulas
(\r{TrLnH}), (\r{TrLnM}) and (\r{TrLnC}) and of the final result
(\r{Ga}), the intermediate calculations were
huge and not only because of the size of the expressions for the
blocks (where {\it Mathematica} \cite{Wolfram}  and its special
package xTensor \cite{xact} for symbolic algebra manipulations
provided a useful assistance),  but also due to the complexity of
all the steps starting from the quadratic expansions. The
correctness of the calculations
has been checked in several ways, as briefly described below.

First, the expression for the operator of the second variational
derivative of the action was verified. The divergence of the
second variational derivative operator (Hessian) with respect to
gravitational fluctuations $h_{\mu\nu}$, from each covariant term
in the gravitational action is separately zero, namely
\beq
\nabla_{\mu}\left(\frac{\delta^{2}S_{{\rm grav},i}}{\delta h_{\mu\nu}\delta h_{\rho\sigma}}\right)=0+O\left(\nabla^{k}{\cal R}^{l},k+2l>4\right),
\quad {\rm where}\quad S_{{\rm QG}}=\sum_i S_{{\rm grav},i}\,.
\eeq
This formula was explicitly checked to the order quadratic in curvatures
and up to total of four derivatives.

Another verification is that, in the total divergent effective action,
we checked a complete cancellation  of terms with singularities in
$y=4\omega_C-6\omega_R$  variable, namely of terms proportional to
$\frac{1}{y}$ and $\frac{1}{y^{2}}$.  This is a nontrivial
cancellation, as the corresponding terms emerge in both
${\rm Tr}\log \hat C$ and ${\rm Tr}\log \hat H$.

Finally, in order to control our {\it Mathematica} \cite{Wolfram} code,
the technically similar computation in the four-derivative QG was
performed. We easily were able to reproduce results about one-loop
UV divergences there \cite{avbar86,Gauss}. We found a complete
agreement for all the coefficients and the same dependence on the
parameter $x$ as in the original papers. This calculation was the
last of the tests.

There is one interesting detail concerning the expression for the
divergences (\ref{Ga}) in the minimal six-derivative theory. It is
easy to note that the coefficient of the $C^2$ term has a general
form
\beq
A_{0}+A_{1}\,x,
\label{eq: sexticdivs}
\eeq
with $x$ defined in (\ref{x}) and $A_0$ and $A_1$ constants.
On the other hand, the other counterterms (i.e., $R^2$ and $E_4$)
come with constant coefficients, independent of $x$. This pattern
is different (one can even say opposite) from the situation in the
four-derivative QG  \cite{frts82,avbar86,Gauss}, where only the
divergence proportional to $R^2$ depends on the analog of (\ref{x}),
i.e., on $x_{\rm 4-der}=\frac{\theta_C}{\theta_R}$, according to
the general form
$B_{-2}\,x_{\rm 4-der}^{-2} +B_{-1} \,x_{\rm 4-der}^{-1} +B_{0}$.
At the same time, e.g., the $C^2$ divergence in the four-derivative
QG comes with the universal (and gauge-fixing independent \cite{a})
coefficient. The mentioned difference is noticeable, regardless we
cannot explain it using some general principles.

One could also analyze a special value of the fundamental ratio
$x$ of the minimal six-derivative theory (\ref{L}), which makes the $C^2$ sector
of UV divergences completely finite. This value is
\beq
x=-\frac{3573}{80}=-44.6625.
\label{special x}
\eeq
We remind the reader for comparison that in the case of quadratic gravity,
the special values for $x_{\rm 4-der}$, which made the coefficient of the
$R^2$ term in divergences vanish, is $x_{\rm 4-der}=3 (3 \pm \sqrt {7})$
\cite{avbar86}.

For the sake of completeness we also remind the reader here of the divergent contributions
to the quantum effective action with the Einstein-Hilbert \cite{SRQG-beta} and the
cosmological constant operators \cite{highderi},
\beq
\Ga^{(1) {\rm EH} , \Lambda}_{\rm div}
&=&
 \frac{1}{2} \frac{\log L^2}{(4\pi)^{2}}\!\int\! d^4x \sqrt{|g|}
\left\{
\frac{5 \theta_C}{6 \omega_C} + \frac{\theta_R}{2 \omega_R}
-  \frac{5 \theta_R}{ \omega_C}
\right\}R
\nonumber
\\
&&
- \frac{1}{2} \frac{\log L^2}{(4\pi)^{2}}\!\int\! d^4x \sqrt{|g|}
\left\{
\frac{5 \omega_\kappa}{2 \omega_C}
- \frac{\omega_\kappa}{6 \omega_R}
- \frac{5}{2} \left( \frac{\theta_C}{\omega_C}  \right)^2
- \frac{1}{2} \left( \frac{\theta_R}{\omega_R}  \right)^2
\right\}.
 \label{GEHL}
\eeq
The sum of (\ref{GEHL}) and  (\ref{Ga}) is the full set of the
relevant one-loop divergences in the six-derivative model
(\ref{L}). The remaining counterterm $\square R$ being
a total derivative is, in fact,
the most difficult to calculate. Also, this makes not much sense to
derive this term in the nonconformal theory, because it is equivalent
to a finite $R^2$-type contribution and the overall  $R^2$ term is
the subject of renormalization conditions.

\section{Beta functions and  \RG}
\label{s4}

Starting from the divergences derived in the previous section,
we can now derive the beta functions of the theory.
From  (\ref{Ga}) and (\ref{GEHL}), the total expression for
the divergent part of the effective action can be presented in the
form
\beq
&&
\Ga^{(1)  }_{\rm ct}
= - \Ga^{(1)  }_{\rm div}
= \frac{1}{2 \epsilon} \frac{1}{(4\pi)^{2}}\!\int\! d^4x \sqrt{|g|}
\Bigg\{
\Big(\frac{2x}{9}+\frac{397}{40}\Big)C^2
- \frac{7}{36}R^{2}
+ \frac{1387}{180}E_4
  \nonumber
  \\
 &&
 \qquad
  \qquad
  \qquad
 -\,
\left[\frac{5 \theta_C}{6 \omega_C} + \frac{\theta_R}{2 \omega_R}
-  \frac{5 \theta_R}{ \omega_C}
 \right]\!R \nonumber \\
 &&
  \qquad
  \qquad
  \qquad
 +\,
\frac{5 \omega_\kappa}{2 \omega_C}
- \frac{\omega_\kappa}{6 \omega_R}
- \frac{5}{2} \left( \frac{\theta_C}{\omega_C}  \right)^2
- \frac{1}{2} \left( \frac{\theta_R}{\omega_R}  \right)^2
 \Bigg\}
\nonumber
\\
&&
  \qquad
  \qquad
=\,\frac{1}{2 \epsilon} \!\int\! d^4x \sqrt{|g|} \left\{
\beta_C C^2 + \beta_R R^2 +\beta_{\rm GB} E_4 + \beta_G R + \beta_\Lambda
\right\}.
\label{formula42}
\eeq
The derivation of the beta functions is a standard operation,
explained, e.g., in the recent textbook \cite{OUP}. Thus, we
give only the final result. The running of all the relevant parameters
in the six-derivative QG is described by the following
renormalization group equations:
\beq
&&
\be_C \,=\, \mu\frac{d \th_C}{d \mu}
 = \frac{1}{(4\pi)^{2}} \Big( \frac{2x}{9}+\frac{397}{40} \Big),
\label{betaW}
\\
&&
\be_R \,=\, \mu\frac{d \th_{R}}{d \mu}
= -  \frac{1}{(4\pi)^{2}} \,\frac{7}{36} \, ,
\label{betaR2}
\\
&&
\be_{\rm GB} \,=\, \mu\frac{d \th_{\rm GB}}{d \mu}
= \frac{1}{(4\pi)^{2}} \,\frac{1387}{180} \, ,
\label{betaGB}
\\
&&
\be_{\kappa} \,=\, \mu\frac{d \omega_{\kappa}}{d \mu}
= - \frac{1}{(4\pi)^{2}} \,
\left[ \frac{5 \theta_C}{6 \omega_C} + \frac{\theta_R}{2 \omega_R}
-  \frac{5 \theta_R}{ \omega_C}
\right]
\label{betaR} \, ,
\\
&&
\be_{\Lambda} \,=\, \mu\frac{d \omega_{\Lambda}}{d \mu}
=  \frac{1}{(4\pi)^{2}} \,
\left[\frac{5 \omega_\kappa}{2 \omega_C}
- \frac{\omega_\kappa}{6 \omega_R}
- \frac{5}{2} \left( \frac{\theta_C}{\omega_C}  \right)^2
- \frac{1}{2} \left( \frac{\theta_R}{\omega_R}  \right)^2
\right]  \, .
\label{betaLambda}
\eeq

It is worth noting that the one-loop beta functions for four-derivative terms
($C^2$, $R^2$, $E_4$) listed above
are exact and universal in the theory (\ref{L}), as we have
demonstrated in Sec.~\r{s2}, while the beta functions for $\omega_\kappa$ and $\omega_\Lambda$
receive also corrections at two-loop and two- and three-loop orders respectively.
 However, all these beta functions (except the $\beta_\Lambda$) can be
modified as the result of  inclusion of additional terms into the initial action of the
theory. For example, this happens if we add terms that are cubic in
Riemann tensor (so called killers). These terms do not spoil the super-renormalizability
of the theory. In particular, adding the term (which can be shown to
be a sum of total derivatives and $O\left({\cal R}^3\right)$ terms
\cite{highderi})
\beq
{\rm GB}_1 \,=\,
R_{\mu\nu\!\rho\si}\square R^{\mu\nu\!\rho\si}
- 4 R_{\mu\nu}\square  R^{\mu\nu} + R\square  R
\label{GB1}
\eeq
is supposed to affect the  \RG  \ equations (\ref{betaW}),
(\ref{betaR2}), (\ref{betaGB}) and  (\ref{betaR}). For example, as computed in
 \cite{SRQG-beta}, the beta function for the parameter $\omega_\kappa$ with the additional contribution of the
 $\om_{\rm GB} \cdot{\rm GB}_1$ term in the Lagrangian (\ref{L}) reads
\beq
\be_{\kappa} \,
=\, - \frac{1}{(4\pi)^{2}} \,
\left[ \frac{5 \theta_C}{6 \omega_C} + \frac{\theta_R}{2 \omega_R}
-  \frac{5 \theta_R}{ \omega_C}
+ \left(\frac{5 \theta_C}{6\omega_C^2}
-  \frac{\theta_R}{ 18 \omega_R^2}  \right) \om_{\rm GB} \right].
\label{betakappagen}
\eeq
Moreover, in this case, it is expected that  beta functions
 $\beta_C$, $\beta_R$ and $\beta_{\rm GB}$ exhibit up to
 quadratic dependence on $\omega_{\rm GB}$. In this section,
we do not introduce this or other similar structures and consider
the quantum theory based on the minimal model (\ref{L}).

In the remaining part of this section, we solve the \RG equations (\ref{betaW}),
(\ref{betaR2}) and  (\ref{betaGB}) for the four-derivative terms,
while the equations for the Newton constant and the cosmological
constant will be considered in the next section for the special case
of Lee-Wick quantum gravity.

Let us start with (\ref{betaW}). In this case the solution is
\beq
\theta_C(t)
\,=\, \theta_C(0) + \beta_C \, t = \theta_C(0) + \frac{1}{(4\pi)^{2}} \Big(\frac{2x}{9}+\frac{397}{40}\Big) \, t \,,
\label{SolW}
\eeq
where the condensed notation with the logarithmic RG-time $t=\log(\mu/\mu_0)$ was used. In the
four-derivative theory, the perturbative coupling constant is $\la_C$ (see, e.g., \cite{OUP}), where
 $\la_C = - \frac{1}{2\th_C}$. In this case, the
stability of the theory requires $\la_C>0$, so $\th_C(t)<0$. Moreover, to have asymptotic
freedom in the UV regime, the signs of $\theta_C(0)$ and $\beta_C$ should be the
same.

It is worth noting that this situation is common for the higher
derivative models since in this case there are, typically,
different degrees of freedom with different masses \cite{frts82}.
When these particles are separated by using auxiliary
fields (see, e.g., \cite{ABSh1} or \cite{OUP} for a simplified
example), the constants of rescaling for the propagators tend to
zero in the case of asymptotic freedom for mutual interactions
between these different degrees of freedom.

The solution to Eq. (\ref{betaR2}) reads
\beq
\theta_R(t) = \theta_R(0) + \beta_R \, t
= \theta_R(0) -   \frac{1}{(4\pi)^{2}} \,\frac{7}{36} \, t \, .
\label{SolR}
\eeq
Similarly to the Weyl-squared term, in a purely four-derivative
theory this would be related to the coupling $\xi$ by
$\theta_R = - \frac{1}{\xi}$ and the stability requires $\xi>0$. In
this case, asymptotic freedom requires $\theta_R(0)$ to be
negative, something that cannot be easily established in the present case.

Finally, the solution of (\ref{betaGB}) is
\beq
\theta_{\rm GB}(t) = \theta_{\rm GB} (0) + \beta_{\rm GB} \, t
= \theta_{\rm GB}(0) + \frac{1}{(4\pi)^{2}} \,\frac{1387}{180}  \, t \, .
\label{SolGB}
\eeq
In the same sense as described above, for the asymptotic freedom
we need $\theta_{\rm GB}(0) > 0$.

Let us note that one can attribute a physical meaning to the running
described above only in the UV (high energy regime) where the
MS-based \RG reflects the physical running of effective charges that
depend on the energy scale. At lower energies the logarithmic running
does not take place due to the Appelquist-Carazzone decoupling
theorem \cite{AC}. For higher derivative QG this theorem has not yet
been tested, but at the semiclassical level there are solid
results in this respect \cite{apco,CodZan} obtained by means of
diagrams and by the extended Schwinger-DeWitt technique
\cite{Avramidi-95,bavi90}. Taking this into account, it is
reasonable to assume that the MS-scheme-based running in
(\ref{SolW}), (\ref{SolR}) and (\ref{SolGB}) takes place for the
energy scales above a threshold defined by the magnitudes of the
masses of the higher derivatives modes, which are actual degrees of freedom (i.e.,
beyond the massless graviton) that are present in the theory.

Indeed, a sufficiently intensive running can change the value of a
relevant parameter even on a restricted interval of energy scales.
In this respect, the most interesting is the running (\ref{SolR}) of
the coefficient $\theta_R$ of the term $R^2$ in the action. However,
it is easy to see that, according to the solution (\r{SolR}), the
running of the coefficient $\theta_R$ cannot provide its value of
the order $10^8\,-\,10^9$, required for the phenomenologically
successful Starobinsky model of inflation \cite{star,star83}. The
perturbative running in super-renormalizable QG is logarithmic and
the beta function is a parameter-independent constant. Thus, exactly
as in the case of other quantum corrections (see, e.g.,
\cite{StabInstab} for the review and further references), this
running cannot change the value of the parameter $\th_{R}$ by many
orders of magnitude starting, e.g., from the value of order one.

\section{Six-derivative Lee-Wick Quantum Gravity}
\label{s5}

The theory proposed and studied by Stelle in \cite{Stelle} shows
good quantum properties like renormalizability and asymptotic
freedom \cite{frts82,avbar86}, but the presence of a
ghost instability at the classical level makes the theory nonunitary in
its original quantization based on the Feynman prescription
\cite{Stelle}. However, recently a new quantum prescription, based
on the Cutkosky, Landshoff, Olive, and Polkinghorne
\cite{CLOP} approach to the Lee-Wick theories \cite{LW1, LW2}, was
introduced by Anselmi, and Piva \cite{AnselmiPiva1, AnselmiPiva2}.
This new prescription allows one to tame the ghost instability of
Stelle's theory. In this way, the unitarity problem is definitely
solved at any perturbative order in the loop expansion
\cite{AnselmiPiva3}. At the classical level the ghost (what Anselmi
and Piva called ``fakeon'' because it can only appear as a virtual
particle) is removed from the spectrum obtained by solving the equations of
motion for the fake field by means of advanced and retarded
Green's functions and by fixing to zero the homogeneous solution
\cite{ClassicalPrescription1, ClassicalPrescription2}. This is the
classical equivalent of removal of ghosts in the quantum
theory from the spectrum of allowed asymptotic states.

The described prescription is very general and can be applied to
real as well as complex poles implying ghosts (fakeons) but also to normal particles.
In particular, it can be applied to make perturbatively unitary the
theory proposed in \cite{LM-Sh, ModestoLeeWick} and named
``Lee-Wick quantum gravity''. This class of theories is based on
the general higher-derivative action proposed in \cite{highderi}
(with more than four derivatives).
Thus, we arrive at a large class of super-renormalizable or finite
and unitary higher-derivatives theories of quantum gravity. In order
to guarantee tree-level unitarity, the theory in
\cite{LM-Sh, ModestoLeeWick} has been designed to possess
only complex conjugate massive poles in the propagator, besides
the massless graviton.
It is worth mentioning that Stelle's theory \cite{Stelle} with the
Anselmi-Piva prescription \cite{AnselmiPiva1, AnselmiPiva2} is the
only strictly renormalizable theory of gravity. On the other hand,
the theories proposed in \cite{highderi,LM-Sh} represent an infinite
class of  super-renormalizable or finite models for QG.  An important
issue of uniqueness of these theories will be addressed in a possible
future project.

The minimal six-derivative Lee-Wick model is obtained starting
from (\ref{6model}) and (\ref{L}), by fixing the parameters of the
action to provide a suitable spectrum of the theory, i.e., one has
to ensure the presence of the usual real graviton field and a pair
of ghosts with complex conjugate poles, with the complex square of the
mass. The action of the theory, including the cosmological constant
term, can be written in the form
\beq
S & = &
\!\int\! d^4x \sqrt{|g|}\,
 \omega_\kappa \left\{  \frac{\omega_\Lambda}{\omega_\kappa}+R
  +  C_{\mu\nu\!\rho\si} \Big( \frac{\theta_{C}}{\omega_\kappa}
 + \frac{\omega_{C}}{\omega_\kappa} \square \Big) C^{\mu\nu\!\rho\si}
 +  R \Big( \frac{\theta_{R}}{\omega_\kappa}
 + \frac{\omega_{R}}{\omega_\kappa} \square \Big) R
  +  \frac{\th_{\rm GB}}{\omega_\kappa} E_4
+\frac{ \om_{\rm GB}}{\omega_\kappa} {\rm GB}_1 \right\},
\label{L2}
\eeq
where we also added for our convenience a generalized Gauss-Bonnet term ${\rm GB}_1$ from (\ref{GB1}), which will also be useful
later in this section. Due to this addition the model ceases to be minimal.

Let us start by considering the classical case without cosmological
constant and show that the spectrum of such a theory contains
only complex conjugate poles and the graviton. Since $E_4$ and
${\rm GB}_1$ terms do not contribute to quadratized action for the graviton
field in Minkowski spacetime (so we set $\omega_\Lambda=0$), the propagator, up to gauge-dependent
terms, in Fourier space and in four spacetime dimensions, reads
\beq
\Pi(k^2) =-\frac{2i}{\om_\ka} \left(\frac{P_2}{H_2(k^2) k^2} - \frac{P_0}{2 H_0(k^2) k^2} \right) \, ,
\eeq
where $P_2$ and $P_0$ are spin projectors (see, e.g., \cite{OUP}) and
\beq
&&
H_2 = 1 - 2 k^2 \frac{\theta_C}{\omega_\kappa}
+ 2k^4 \frac{\omega_C}{\omega_\kappa} ,
\label{H2}
\\
&&
H_0 = 1+ 6k^2 \frac{\theta_R}{\omega_\kappa}
- 6 k^4 \frac{\omega_R}{\omega_\kappa} .
\label{H0}
\eeq
We remark that the above form of the coefficient in front of
the spin-0 part, proportional to the projector $P_0$, of~the~propagator, although originally it was obtained
in the harmonic gauge, is valid generally for any gauge choice.
Actually, functions $H_2(k^2)$ and $H_0(k^2)$ do not show any
dependence on gauge-fixing parameters. Hence also masses
of additional spin-0 or spin-2 modes are completely gauge-fixing-independent \cite{OUP}, both on the flat spacetime as
well as around (A)dS backgrounds.

If we define $z=k^2$, the location of the zeros for both
$H_2$ and $H_0$ is
\beq
&&
H_2(z) = 0 \quad \Longrightarrow \quad
z = \frac{\theta_C \pm \sqrt{\theta_C^2
- 2 \omega_C \omega_\kappa}}{2 \omega_C} ,
\label{z2}
\\
&&
H_0(z) = 0 \quad \Longrightarrow \quad z
=  \frac{3 \theta_R \pm \sqrt{3} \sqrt{3 \theta_R^2
+ 2  \omega_R \omega_\kappa }}{6 \omega_R} .
\label{z0}
\eeq
Thus, the conditions to have two pairs of complex conjugate poles are
\beq
&&  \theta_C^2 - 2 \omega_C \omega_\kappa < 0 \, ,
\label{C1} \\
&&
3 \theta_R^2 + 2  \omega_R \omega_\kappa < 0  \, .
\label{C2}
\eeq
It is evident from the above constraints (\ref{C1}) and (\ref{C2}),
that for $\omega_\kappa> 0$, we have to assume $ \omega_R <0$ (below we will use $\om_R=-|\om_R|$)
and $\omega_C> 0$.

At the quantum level, the relations (\ref{C1}) and (\ref{C2}) may
change because of the running with the energy scale $\mu$   and some coupling constants like $\theta_C$, $\theta_R$, $\omega_\kappa$  above, acquire nontrivial $t$-dependence.
 We remark that
in an hypothetical ideal situation
one should look for the full answer in the full quantum effective action and from this
functional,
one should read the positions and residues
of the poles of the propagator. These would then correspond to the effects of true quantum
propagation of dressed particles.
The effective action functional $\Gamma$ for this task should be understood on the tree-level.
However, as it is well known the computation
of the effective action is a very difficult task. Even at the one-loop level all its finite terms are much
more complicated than just the technically demanding calculation we present in this paper.
Therefore, our strategy is to look at the RG-improved form of the tree-level action of the theory.
 In this approach, we substitute tree-level classical values of
couplings by their quantum running analogs obtained by integrating the
beta functions that we have computed in the first part of the paper.
This constitutes a first step towards taking the impact of quantum effects on the
propagation of modes and on the spectral properties of the quantum theory since we do not have at the moment the knowledge of the full effective action and presently its full form is beyond our computational reach.

From such an RG-improved classical
action of the theory we derive the propagator and analyze the structure of poles and
their residues for all dynamical particles of the model by standard
procedure (treating the RG-corrected action as given at the tree-level). We also want to
remark that we expect that the quantum
perturbative corrections (for example at the one-loop level) will not
change too much the positions of the
poles compared to the classical values. Hence, according to the perturbative approximation very likely
a classical Lee-Wick theory will also remain such after the inclusion of quantum corrections.

Such an RG-improvement procedure for the action and its couplings is standard in QFT
without gravity and around flat spacetime. We strongly think that the case of gravity, or of
$6$-derivative QG in particular, should not be more special in this respect.
Gravitation shall not be treated like completely exceptional QFT, it is simply a quite peculiar
QFT with a special particle content and a special set of symmetries.
One can easily understand this by viewing gravity from the higher spin perspective.
Models with higher spins are perhaps the most profoundly  generalized QFTs,
while theories of pure QG are special cases of former ones.
Our philosophy is to use and apply the same methods which were already
successfully applied in the case of QFT on a flat background
for nongravitational interactions, but this
time to quantum gravity. We think that gravity is not much different
from other interactions as considered in the particle physics framework
and could be treated on the same footing as they are and could be, for example,
 completely understood in the Feynman
way of viewing gravity \cite{Feynman} as another spin-2 field (on flat background)
with self-interactions dictated by a suitable gauge principle.

 Furthermore,
in order to verify the stability of the spectrum, we must also consider
the running of the Newton constant $G$ related to the $\omega_\ka$
coupling constant. Hence, we should check
under what conditions on the free parameters $\om_C$, $\om_R$,
initial conditions $\th_C (0)$, $\th_R (0)$ and $\om_\kappa(0)$, the
ghosts remain forming complex conjugate pairs at any energy scale. Here we set $\om_{\rm GB}=0$ in (\ref{L2}).
Making explicit the dependence on $t$, (\ref{C1}) and (\ref{C2}) become
\beq
&&  \theta_C(t)^2 - 2 \omega_C \, \omega_\kappa(t) < 0 \, ,
\label{C1t}
\\
&&
3 \theta_R(t)^2 -  2  | \omega_R | \, \omega_\kappa(t) < 0  \, .
\label{C2t}
\eeq
Using the conditions discussed in the previous section,
the requirement of having an asymptotically free theory implies
$ \theta_C (0)> 0$ (for $\beta_C > 0$) and $\theta_R (0) <0$ (for $\beta_R < 0$) consistently with  (\ref{SolW}) and (\ref{SolR}) respectively.
Hence, for $\theta_R(0) < 0$ we can rephrase (\ref{SolW}) and (\ref{SolR}) as
\beq
&& \theta_C(t) = \theta_C(0) + \beta_C \, t = \theta_C(0) + \frac{1}{(4\pi)^{2}} \Big( \frac{2 \omega_C}{9 \omega_R}+\frac{397}{40} \Big) \, t,
\label{SolWLW}
\\
&& \theta_R(t) = \theta_R(0) + \beta_R \, t = - \left[ |\theta_R(0)| +   \frac{1}{(4\pi)^{2}} \,\frac{7}{36} \, t \right].
\label{SolRLR}
\eeq
Furthermore, if we assume $\theta_C(0) > 0$, we need the following condition to be satisfied to ensure asymptotic freedom:
\beq
\frac{397}{40}- \frac{2 \omega_C}{9 |\omega_R| } > 0 \quad \Longrightarrow \quad \frac{\omega_C}{|\omega_R| } < \frac{ 3573}{80} \, .
\eeq

Finally and most importantly, we should ensure that
there are no tachyons in the spectrum, namely that the real parts of
the two roots in (\ref{z2}) and (\ref{z0}) are negative according to
our signature of the metric. Unfortunately, this is not possible,
because in (\ref{z2}) $\theta_C(t)$ and $\omega_C$ are both
positive as well as in (\ref{z0}) $\theta_R(t)$ and $\omega_R$ are both
negative. In the spin-zero sector, we could flip the sign of $\omega_R$, but in this case
due to (\ref{C2}), the ghost would become a real particle  (see, e.g., \cite{Yamamoto1970} for
a detailed discussion\footnote{According to  \cite{Yamamoto1970}, in order to have a theory that satisfies macrocausality the group velocity must be equal or smaller than the speed of light and, hence, the real part of the mass square parameter must be positive or zero.}). This problem cannot be solved in the
framework of the minimal model  (\ref{L}). Thus,
the only way out is to extend the Lagrangian (\ref{L}) by adding
new operators, like ${\rm GB}_1$ in (\ref{L2}), cubic in the Riemann tensor and/or its contractions,  to
make the beta function of the $\th_R$ coupling positive. In the latter case, we can achieve asymptotic
freedom and condition of no tachyons at the same time, because (\ref{SolR})
will be replaced by a new running with a positive beta function $\tilde{\be}_R>0$ and with a positive initial condition $\th_R(0)>0$, namely
\beq
 \theta_R(t) = \theta_R(0) + \tilde{\beta}_R \, t =  |\theta_R(0)| +   \tilde{\beta}_R \, t  \,
\label{SolRLR-a}
\eeq
with a positive overall sign.
Much simpler is to eliminate the tachyon for the spin-two sector. As
stated above, in (\ref{C1}) $\omega_C$ cannot be negative if we want
to have a complex pole. However, we can fix $x$ to make zero the
beta function $\beta_C$ or to flip the overall sign in (\ref{SolW})
and choose, at the same time, $\theta_C(0)$ to be negative.
In this case, we guarantee that $\theta_C(t)<0$, so the real
part of the mass square parameter in (\ref{z2}) is negative.
Also, under these conditions, the group velocity is smaller than the speed of light \cite{Yamamoto1970} and macrocausality is satisfied \cite{Giaccari:2018nzr}.

In the six-derivative theory, the cosmological constant $\omega_\Lambda$ runs with
the energy and the assumption of being in Minkowski spacetime
is not valid anymore. Therefore, we must study the stability of the
theory in (A)dS  backgrounds. To this end, we shall follow the general analysis
in \cite{ModestoAdS, Allen,meBiswasMaz,meBiswasMazLONG}.
At the quantum level, the conditions (\ref{C1t}) and (\ref{C2t}) are
not sufficient to guarantee the absence of ghosts and we have
to study the perturbative spectrum of the theory in (A)dS.

In order to expand the action (\ref{L2}) to the second order, in
the graviton fluctuation $h_{\mu\nu}$, around the onshell (A)dS background
($\bar{g}_{\mu\nu}$), we use the following York decomposition of the
graviton field \cite{ModestoAdS,Percacci}:
\beq
&&
g_{\mu\nu} = \bar{g}_{\mu\nu} + h_{\mu\nu} \, ,
\nonumber
\\
&&
h_{\mu\nu} = h_{\mu\nu}^{\bot} + \nabla_{(\mu} A^\bot_{\nu)}
+ \left( \nabla_\mu \nabla_\nu
- \frac{1}{4} g_{\mu\nu} \square \right) B + \frac{1}{4} g_{\mu\nu} h\,,
\eeq
where the spin-two transverse and traceless fluctuation $h_{\mu\nu}^{\bot}$ contains $5$
degrees of freedom because it satisfies
$\nabla^\mu  h_{\mu\nu}^{\bot} = g^{\mu\nu} h_{\mu\nu}^{\bot} =0$.
The transverse vector $A^\bot_{\nu}$,  satisfying
$\nabla^\nu A^\bot_{\nu}=0$, accounts for $3$ degrees of
freedom. Finally, $B$ and $h$ are two real scalars. However,
$A^\bot_{\nu}$ automatically drops out of the second variation
of the action of the theory (\ref{L2}) due to the diffeomorphism invariance and out of the two scalars only the following combination
$\phi = \square B - h$ appears there. Here, again, we set $\om_{\rm GB}=0$ in (\ref{L2}). We end up with the following second
order variation of the action around an (A)dS background parametrized by the cosmological constant
$\Lambda$ (according to the relation $\bar{R}_{\mu\nu}=\Lambda \bar{g}_{\mu\nu}$) \cite{ModestoAdS,Allen,meBiswasMaz,meBiswasMazLONG},
\beq
 S^{(2)}_{\rm (A)dS} = \frac{1}{2} \!\int\! d^4 x \sqrt{| \bar{g}|}
 \left\{ \tilde{h}^{\bot \mu\nu}
\left( \square - \frac{2}{3} \Lambda
\right)  H_2(\square, \Lambda) \tilde{h}^\bot_{\mu\nu}
- \tilde{\phi} \left( \square + \frac{4}{3} \Lambda
\right)  H_0(\square, \Lambda) \tilde{\phi}
\right\},
\label{variationADS}
\eeq
where we introduced the following two polynomials (being analogs of (\ref{H2}) and (\ref{H0}) in the case $\Lambda\neq0$),
\beq
&&
H_2(\square, \Lambda)
= 1 + 8 \Lambda \frac{\theta_R}{\omega_\kappa}
+ 2 \left( \square -  \frac{4}{3} \Lambda \right)
\left[\frac{\theta_C}{\omega_\kappa}
+ \frac{\omega_C}{\omega_\kappa} \left( \square
+ \frac{4}{3} \Lambda \right) \right] \, ,
\label{H2H2}
\\
&&
H_0(\square, \Lambda)
= 1 + 8 \Lambda \frac{\theta_R}{\omega_\kappa}
- 6  \left( \square + \frac{4}{3} \Lambda \right)
 \left[ \frac{\theta_R}{\omega_\kappa}
 + \frac{\omega_R}{\omega_\kappa} \square \right] \, ,
\label{H2H0}
\eeq
and the canonically normalized fields
\beq
\tilde{h}^{\bot}_{ \mu\nu} = \frac12\, M_{\rm P} \,  {h}^{\bot}_{ \mu\nu},
\qquad
\tilde{\phi} = \sqrt{\frac{3}{32}}\,M_{\rm P} \, \phi,
\qquad
\frac12\,M^2_{\rm P} = \frac{1}{2}\, \kappa_4^{2} = \omega_\kappa.
\eeq
In order to have the same stability properties (at the linear level) in the six-derivative theory as in Einstein's two-derivative gravity, we require the two polynomials
$H_2$ and $H_0$ to have only complex conjugate roots. This requirement corresponds to the Lee-Wick prescription in the presence of the cosmological constant $\Lambda$.
Introducing again the variable $z=- \square = k^2 $ (here by $\square=\bar{g}^{\mu\nu}\nabla_\mu\nabla_\nu$ we mean the d'Alembert operator in the (A)dS spacetime), the roots of (\ref{H2H2}) and (\ref{H2H0}) are
\beq
&& H_2(z) = 0 \quad \Longrightarrow \quad z = \frac{3 \theta_C \pm \sqrt{\Delta_2}}{6 \omega_C} \, , \label{H2z}\\
&&
H_0(z) = 0 \quad \Longrightarrow \quad z =  \frac{3 \theta_R + 4 \Lambda \omega_R \pm  \sqrt{ \Delta_0 }}{6 \omega_R} \label{H0z}\, ,
\eeq
where
\beq
&& \Delta_2 = ( 3 \theta_C + 8 \Lambda \omega_C)^2 - 18 \omega_C ( \omega_\kappa + 8 \Lambda \theta_R ) \, ,\label{Delta200}
\\
&& \Delta_0= 9 \theta_R^2 + 24 \Lambda \theta_R \omega_R + 2  \omega_R
 ( 3 \omega_\kappa + 8 \Lambda^2 \om_R ) \, .\label{Delta000}
\eeq
Now we assume $\omega_R < 0$ and $\omega_C>0$ according to the
conditions in (\ref{C1}) and (\ref{C2}) in Minkowski spacetime. Moreover, we take
$\theta_R(0) < 0$, in order to achieve asymptotic freedom in the UV regime. The running onshell
cosmological constant parameter $\Lambda(t)$ is defined by (when we use effective equations of motion from Einstein's gravity)
\beq
 \Lambda(t) = - \frac{\omega_\Lambda(t)}{2\omega_\kappa(t)} \, .
 \label{cosmoomega}
 \eeq
In the full quantum theory, for consistency we end up with the situation that the scalar curvature
 of the onshell (A)dS background upon which quantum fluctuations are considered,
 is proportional to the running cosmological constant parameter $\Lambda(t)$ according
 to the relation $\bar{R}=4\Lambda(t)$. This signifies that the background curvature
 depends on the energy scale $\mu$ used to probe it by means of fluctuations.
 For consistency, the classical conditions for Lee-Wick particles in (A)dS $\Delta_2<0$ and
 $\Delta_0<0$ with (\ref{Delta200}) and (\ref{Delta000}) have now to be considered
 with the running parameter $\Lambda(t)$ instead of $\Lambda$.
Therefore,
the conditions (\ref{C1t}) and (\ref{C2t}) are now replaced by
 \beq
&&
\big[ 3 \theta_C(t) + 8 \Lambda(t) \omega_C\big]^2
- 18 \omega_C \big[ \omega_\kappa(t) + 8 \Lambda(t) \theta_R(t) \big] < 0 \, ,
\label{C1L}
\\
&&
9 \theta_R(t)^2 + 24 \Lambda(t) \theta_R(t) \omega_R
+ 2  \omega_R \big[ 3 \omega_\kappa(t)
+ 8 \Lambda(t)^2 \om_R \big]  < 0  \, .
\label{C2L}
\eeq
In the case of the modified beta function for $\om_R$ as in ({\ref{SolRLR-a}), it proves useful to make the replacement
$\theta_R(0) \to - \theta_R(0)$. Thus, $\theta_R(0)$ must be chosen positive
to achieve the asymptotic freedom in agreement with ({\ref{SolRLR-a}).

Integrating the \RG equations (\ref{betaR}) and (\ref{betaLambda}),
we finally get the following general solutions:
\beq
& \omega_\kappa (t) = \omega_\kappa &-\frac{ 3  \theta _R \omega _C+5 \theta _C \omega _R-30\, \theta _R
   \omega _R}{96\, \pi ^2\, \omega _C \omega
   _R}t-\frac{ 38\, \omega _C+3993\, \omega _R}{221184\,
   \pi ^4\, \omega _C \omega _R}t^2,\label{solkappa}\\
&\omega_\Lambda(t) = \omega_\Lambda& -\frac{ \omega _R \left(15\, \theta _C^2 \omega _R-15\,
   \omega _C \omega _R \omega _{\kappa }+\omega _C^2 \omega _{\kappa
   }\right)+3 \theta _R^2 \omega _C^2 }{96 \,\pi ^2\, \omega
   _C^2 \omega _R^2}t\nn\\
   &&\hspace{-0.cm}+\frac{10\, \theta _R \left(-15\,
   \omega _C \omega _R+2 \omega _C^2+90\, \omega _R^2\right)-\theta _C
   \omega _R \left(70\, \omega _C+3723\, \omega _R\right)}{36864\, \pi
   ^4\, \omega _C^2 \omega _R^2}t^2\nn\\
   &&\hspace{-0.cm} -\frac{ 537450\, \omega _C \omega
   _R+7000\, \omega _C^2+13365279\, \omega _R^2}{637009920\,\pi
   ^6\, \omega _C^2 \omega _R^2}t^3,
\label{solLambda000}
\eeq
where $\theta_C, \theta_R$ stand for the initial values
$\theta_C(0), \theta_R(0)$.
The physical (running) cosmological constant given in (\ref{cosmoomega}) is proportional to the ratio of $\omega_\Lambda(t)$ and $\omega_\kappa(t)$, which
are   cubic and  quadratic  polynomials of $t$ respectively, namely we have
\beq
&& \omega_\kappa(t) = \omega_\kappa
 + a_\kappa \, t + b_\kappa \, t^2 \, ,
\label{runningkappa}\\
&&  \omega_\Lambda(t) = \omega_\Lambda
+ a_\Lambda \, t + b_\Lambda \, t^2 + c_\Lambda \, t^3 \, ,\label{runningLambda}\eeq
 where the above coefficients $a_\kappa$, $b_\kappa$, $a_\Lambda$, $b_\Lambda$ and $c_\Lambda$ are just nonrunning constants and
 $\omega_\kappa$ and  $\omega_\Lambda$ are initial values (at $t=0$) of running couplings $\omega_\kappa(t)$ and $\omega_\Lambda(t)$ respectively. Therefore, in general,  the
 cosmological constant grows linearly in the ultraviolet regime, i.e.
\beq
\Lambda(t) \sim t.\label{runningLambda00}
\eeq
However, the space of parameters for couplings in front
 of cubic operators, like $\omega_{\rm GB}$, is certainly large enough
to make vanish the linear contribution in $t$ to the
cosmological constant $\Lambda(t)$ that now can approach a constant value
in the ultraviolet regime.
 One can see that this is not possible just with playing with values of
$\om_C$  and $\om_R$ alone since the trinomial
\beq537450\, \omega _C \omega
   _R+7000\, \omega _C^2+13365279\, \omega _R^2
\eeq
does not have real solutions (its discriminant is negative).
Note that the two parameters $\omega_R$ and $\omega_C$ do
not scale with the energy, i.e. they do not run.

One can envisage three different cases for the UV-limiting behavior
of the physical cosmological constant. In the first (worst) scenario,
the term proportional to $t^3$ in (\ref{solLambda000}) is not canceled and the $\Lambda(t)$ goes
to infinity in the UV regime and hence the spacetime becomes fuzzy
 or with a foamlike structure \cite{avbar86,frts82}. This can be avoided by inclusion
 of cubic operators with tuned coefficients. In the second case,
the cosmological constant goes to zero in the high energy regime
and we end up with the Minkowski vacuum in the ultraviolet.
In the last scenario, if we do not select the initial conditions
 entailing the second case, we have to face (A)dS vacuum at short distance
 with some finite radius of curvature.

By inclusion of the cubic in curvature operators,
one can imagine a situation when the Lee-Wick conditions (\ref{C1L}) and
(\ref{C2L}) are always satisfied, namely the poles of ghosts stay in a complex pair at any
energy scale from IR to UV. In the second case considered above,
with the initial condition $\Lambda(0) = 0$, running $\Lambda(t)$ starts from zero in the IR,
it increases/decreases and falls again to be zero in the UV. For the third
choice the spacetime evolves from (anti)- de Sitter
in the IR to another (A)dS in the UV. Therefore, in the ultraviolet regime,
we can have the following two sensible scenarios: free gravitons propagating
on Minkowski or free gravitons propagating in (A)dS backgrounds.

It is worthwhile to make a comment about the generality of the
results obtained here. As we mentioned in the previous section,
the operator ${\rm GB}_1$ and other cubic operators in the Riemann tensor and/or
its contractions may affect the beta functions for the couplings
$\theta_C$, $\theta_R$, $\theta_{\rm GB}$ and $\omega_\kappa$. In the present work,
we  did not compute these contribution of ${\rm GB}_1$.
However, the  contributions to $\beta_C$, $\beta_R$ and $\beta_{\rm GB}$  dependent on
$\om_{\rm GB}$
can affect the analysis in this section. Indeed, the
$\om_{\rm GB}$-dependent contributions can flip the signs of
 all these beta functions and the corresponding conditions for
 asymptotic freedom in respective couplings $\th_C$, $\th_R$ and $\th_{\rm GB}$
  will require to flip the signs of $\theta_C(0)$, $\theta_R(0)$ and $\theta_{\rm GB}(0)$ respectively as well.

The RG-improved version of the action is a reasonable approximation
to solve the problem of the locations of poles of the propagator and their shifts due to quantum effects.
It is definitely true that conditions for complex conjugate pairs of poles for LW particles here are
rather restricting. However, since we expect that quantum corrections are small and perturbative, then we
expect the same situation to be present on the full quantum level and we express the hope that it is
probably possible to remain with the LW QG also on the one-loop level when quantum perturbative effects
are taken fully into account.

\section{Asymptotic freedom}
\label{s6}

In this section, we discuss the issue and necessary conditions for asymptotic freedom (AF) of the six-derivative QG models.
First, we remark that usual textbook definitions of
AF as typically used in two-derivative theories and for couplings
like Yang-Mills charge $g$ (that is requiring that $\beta_g<0$)
could be a bit misleading.
Here one has to exert special care when discussing AF in higher derivative theories.
Actually, we want to point out that in the more general situation the
condition $\beta<0$ loses its validity. For example, even in the example
of QCD, the beta function of the reparametrized coupling $\omega=g^{-2}$ is not negative in
the deep UV regime. Actually near the trivial Gaussian UV fixed point (FP) (still describing AF pure Yang-Mills theory),
it reaches a constant positive asymptotic value $\frac{11}{3} \frac{C_2(G)}{(4\pi)^2}$.
Based on this simple example, we need  to allow for an extension of
the definition of AF for the case of different coupling parametrizations
and also for theories that in the kinetic terms for fluctuations contain
higher than two-derivative terms. In a general setup, one has to be very careful
in using the condition $\beta<0$ for AF and assigning a potential physical meaning to it.

Actually, precisely such need we find
in our six-derivative gravitational theory where we write our beta functions
for $\omega$-like (like in the QCD example discussed above) type of couplings (compare with formula (\ref{L})).
Moreover, in higher derivative theories the couplings have not yet been measured and we can always  consistently achieve AF by
 switching at our wish the signs of the classical coupling constants, which are actually the initial conditions for the RG flow equations.

Therefore, we propose the following definition of physical asymptotic freedom
(PAF). It is simply termed as the requirement that the interaction terms
are suppressed compared to the kinetic terms describing free propagation
of particles with quantum effects included. What would be a more physical
definition of AF? The definition of PAF above  is certainly more physical than the general theoretical
requirement of $\beta<0$ (holding only for some way of writing
couplings and only in two-derivative theories). In more general theories with more general types of couplings,
the system of  beta functions and conjunctions of conditions and inequalities for AF  are more intricate.

Our definition of PAF is one of the simplest and is very physical (and
this implies that quantum scattering amplitudes, of course due to interactions,
should go to zero in the deep UV regime).
The reference for such a definition of PAF is the seminal paper by Fradkin
and  Tseytlin \cite{frts82} and \cite{avbar86}, where it is shown that our accepted definition is likely
the physically reasonable one. We here  follow exactly the same rescaling procedure as done by
the authors above
(see below where we will perform it explicitly and with all the details).
 The procedure is based on the rescaling of the graviton field by the same overall factor
  and showing that the physical interactions are
suppressed compared to kinetic operators and that this suppression factor really
tends to zero asymptotically at infinite energy.

In analogy with Stelle's quadratic gravitational theory \cite{Stelle}, $\omega_\kappa$ in our paper plays the role of the coupling $f^{-2}$ in Stelle's theory and all the other couplings, which can run to a constant value in the UV, run logarithmically with the energy scale, or they do not run at all because they are in front of terms in the action containing more than four derivatives of the metric, playing the role of general $\omega$ couplings in Stelle's theory (see \cite{frts82} about the definitions of $f$ and $\omega$ there). We are here assuming that in the six-derivative theory all the running couplings are AF in the UV according to the common definition, namely there are no Landau poles for them at any finite energy scale.

Additional conditions for the RG flow towards an AF theory in the UV regime are that
the initial values for the couplings, which we set conventionally at $t=0$ are selected in order to avoid tachyons and real ghosts.
This choice must make physical sense for the theory, that is the theory must be well defined and for example in gauge theories we cannot have a negative energy of perturbative excitations around vacuum (this enforces that the electric or YM charge is real, not imaginary).
Moreover, we demand that the flow of all running couplings does not meet any singularity, namely there are no Landau poles for any finite value of the logarithmic energy scale $t$. The zero point values are also excluded from the flow, because there the coupling could change the sign and this could for example invalidate the positivity of energy as discussed above.
These two conditions translate into inequalities for the initial values of the couplings and their first derivatives at $t=0$. For example, assuming that $\omega_C(t=0)>0$ for AF we must also require that $\beta_C(t=0)>0$. (For $\omega$-type couplings it is natural to assume $\beta_\omega>0$ for AF in UV, contrary to $g$-type couplings, like the electric or YM charge, for which we shall assume $\beta_e<0$ or $\beta_g<0$ for AF in the UV regime.)
To clarify we point out that for the $\omega$-type of couplings, the situation with $\omega=0$ corresponds to what is commonly called as Landau pole and $\omega\to\infty$ is a trivial Gaussian fixed point of the RG flow. In terms of $g$-type couplings, the situation is opposite and $g=0$ is a free asymptotically noninteracting theory and $g\to\infty$  is a sign of loss of perturbativity and of meeting a hypothetical Landau pole.

Assuming that all the couplings flow in the regular (described above) way and reach AF in the UV regime, we can eventually perform a rescaling analysis of the graviton field in order to compare the kinetic term with the interactions. For this propose we rescale by the coupling $\omega_\kappa$ to have the two-derivative kinetic term of the graviton field in the canonical form.

Therefore one can easily prove that PAF is the feature of our model in the
UV, provided the specific signs of beta functions that we have computed in earlier sections of this paper.
By such an elegant analysis with field rescalings we do much more that
just analysis of signs of beta functions. And we show below that PAF works
in a more complicated setup of higher
derivative kinetic (this part is usually not considered in standard textbooks on QFTs) and
interaction terms.

In this way we convince  the reader that PAF is a
more physical criterion than theoretical $\beta<0$ and that it  indeed describes the physical features of quantum interactions
compared to quantum propagation of free modes. Moreover, this is also the definition used in Stelle’s theory and higher derivative gauge theory in \cite{frts82} (see the appendix of that paper). Therefore we
confidently can say that under specific conditions
the theory is free in the UV since only propagation of free modes
matters in the UV regime. In short, we find a free graviton particle in the UV.

In order to complete the whole story, here we explicitly show the asymptotic
freedom of the theory (\ref{L2}), assuming the cosmological constant to be zero or constant
in the ultraviolet regime realizing the second case discussed above. Since the
Newton constant proportional to $\om_\ka^{-1}$ in the UV falls off faster than all other running couplings (cf. (\ref{SolW}), (\ref{SolR}), (\ref{SolGB}), (\ref{runningkappa}) and (\ref{runningLambda})),
we rescale the graviton field as follows:
\beq
h_{\mu \nu} =  \frac{1}{\sqrt{\omega_\kappa}}\,\tilde{h}_{\mu\nu},
\qquad
g_{\mu\nu} = \eta_{\mu \nu}
+  \frac{1}{\sqrt{\omega_\kappa}} \tilde{h}_{\mu\nu},
\qquad
[\tilde{h}_{\mu\nu}] = E^1 .
\eeq
The effective (RG-improved) action from (\ref{L2}), at the second or higher order in the perturbation
$\tilde{h}_{\mu\nu}$ and around flat spacetime, reads
\beq
S^{(2)} & = & \!\int\! d^4x
\left.\bigg\{
-2 \Lambda(t) \left( \sqrt{| g|} \right)^{(2)}
  + {O}
  \left( \frac{\Lambda(t)  }{ \sqrt{\omega_\kappa(t)}} \tilde{h}^3 \right)+
\left( \sqrt{| g|} R \right)^{(2)}
+ {O}
\bigg( \frac{ \pa^2\,\tilde{h}^3}{\sqrt{\om_\kappa (t)}}\bigg)
 \right.
 \nonumber \\
 &&
 \hspace{0.4cm}
 +
 \left.
  C^{(1)}_{\mu\nu\!\rho\si} \left( \frac{\theta_{C} (t) }{\omega_\kappa (t)}
 + \frac{\omega_{C}}{\omega_\kappa(t)} \square^{(0)} \right) C^{(1)\mu\nu\!\rho\si}
 + {O} \left( \frac{\theta_{C} (t) }{\om_\ka (t)^{3/2}}
 \, \partial^4 \tilde{h}^3 \right)
   + {O} \left( \frac{\omega_{C}}{\omega_\kappa (t)^{3/2}}
   \, \partial^6 \tilde{h}^3 \right)
 \right.
 \nonumber\\
 &&
 \hspace{0.4cm}
 +
 \left.
 R^{(1)} \left( \frac{\theta_{R} (t) }{\omega_\kappa (t)}
 + \frac{\omega_{R}}{\omega_\kappa(t)} \square^{(0)} \right) R^{(1)}
 + {O} \left( \frac{\theta_{R} (t) }{\omega_\kappa (t)^{3/2}}
 \, \partial^4 \tilde{h}^3 \right)
 + {O} \left( \frac{\omega_{R}}{\omega_\kappa (t)^{3/2} }
 \,  \partial^6 \tilde{h}^3 \right)
 \right.
 \nonumber
 \\
 &&
 \hspace{0.4cm}
 +
 \left.
 \frac{\th_{\rm GB}(t)}{\omega_\kappa(t) } E_4^{(2)}
  + {O} \left( \frac{\theta_{\rm GB} (t) }{\omega_\kappa (t)^{3/2}  }
  \, \partial^4 \tilde{h}^3 \right)
+\frac{ \om_{\rm GB}}{\omega_\kappa(t)} {\rm GB}_1^{(2)}
 + {O} \left( \frac{\om_{\rm GB} }{\omega_\kappa (t)^{3/2}  }
 \, \partial^6 \tilde{h}^3 \right)
  \right\}
  \
  \\
  & = &
   \!\int\! d^4x
  \left\{  -2 \Lambda(t) \left( \sqrt{| g|} \right)^{(2)}
  + \left( \sqrt{| g|} R \right)^{(2)}
+  C^{(1)}_{\mu\nu\!\rho\si} \left( \frac{\theta_{C} (t) }{\omega_\kappa (t)} + \frac{\omega_{C}}{\omega_\kappa(t)} \square^{(0)} \right) C^{(1)\mu\nu\!\rho\si}
 \right.
 \nonumber \\
 &&
 \hspace{0.4cm}
 +
 \left.
 R^{(1)} \left( \frac{\theta_{R} (t) }{\om_\ka (t)}
 + \frac{\omega_{R}}{\omega_\kappa(t)} \square^{(0)} \right) R^{(1)}
  +  \frac{\th_{\rm GB}(t)}{\omega_\kappa(t) } E_4^{(2)}
+\frac{ \om_{\rm GB}}{\omega_\kappa(t)} {\rm GB}_1^{(2)} +O\left(\frac1t{\tilde h}^3\right)
\right\} ,
\label{L3}
\eeq
where we explicitly compared the terms with the same number of
derivatives and we neglect writing terms linear in the perturbation $\tilde{h}_{\mu\nu}$ due to onshell condition.
The above higher order terms are symbolically indicated by
${O}(\ldots)$
and $\square^{(0)}$ means the d'Alembert operator constructed with the
unperturbed metric.
Moreover, with the labels $\,^{(1)}$ and $\,^{(2)}$ we pointed the expansion at the first and second order in the perturbation $\tilde{h}_{\mu\nu}$.
The dominant kinetic terms (quadratic in $\tilde{h}_{\mu\nu}$) around flat background
were all indicated in (\ref{L3}), while interaction terms, possibly containing
also derivatives, are of order $\tilde{h}_{\mu\nu}^3$
or higher. Due to the schematic form of the running in the UV,
\beq
\om_\ka(t)\sim t^2 \quad \implies \quad \sqrt{\om_\ka(t)}\sim t, \quad \th_{C,R,{\rm GB}}(t) \sim t, \quad \quad \om_{C,R,{\rm GB}}={\rm const} \sim t^0,
\label{asrel}
\eeq
one sees that the ratios in $O(\ldots)$ brackets are suppressed
 for large $t$ (e.g., $\th_C(t)/\om_\ka(t)^{3/2} \sim t^{-2}$, $\om_C/\om_\ka(t)^{3/2} \sim t^{-3}$).
 Similarly $1/\sqrt{\om_\ka(t)}\sim t^{-1}$ and we have also assumed,
 consistently with the previous discussion, that $\La(t)\sim t^0$ -- reaches a constant
 value (or zero) in the UV regime. One could also instead of looking at asymptotic relations in (\ref{asrel}) valid for $t\to+\infty$, consider the exact form of the RG flows for couplings as found in (\ref{solkappa}) and in (\ref{SolWLW}), (\ref{SolRLR}).
This is why it is clear from the expansion (\ref{L3})  that
in the ultraviolet regime, $t\gg1$, all interaction terms involving three or
more gravitons are suppressed with respect to the kinetic terms.

One can also see the universal feature of the rescaling procedure that we have applied above.
Indeed, the theory turns out to be AF regardless of which
running coupling parameter is used to do the rescaling of the gravitational fluctuation field
 $h_{\mu\nu}$. Above we used $\omega_\kappa(t)$, provided it has a regular form of RG running and the corresponding inverse coupling $\omega_\kappa^{-1}(t)$ does not meet any zero
($\omega_\kappa^{-1} \to 0$) (giving AF theory), nor a Landau pole ($\omega_\kappa^{-1} \to \infty$) at any finite $t$ value.
 But for this purpose we could equally well use a running parameter $\theta_C(t)$ or $\theta_R(t)$, provided that the same conditions for the regular RG flows are satisfied. We still should assume that the cosmological constant parameter $\Lambda$ does not exhibit any RG running. This universality of the rescaling analysis is another
confirmation that our results about PAF in the UV regime in our theory are very robust.

So far so good, but unfortunately both ghosts in (\ref{H2z}) and in (\ref{H0z}) have
positive real parts and, therefore, are tachyons in our signature of the metric
\cite{Yamamoto1970}.
One of the ways to overcome this obstacle consists in adding
at least two operators cubic in the curvature  to provide the
vanishing beta functions for $\theta_C$ and $\theta_R$.
Alternatively, the beta function $\beta_C$ can be made zero also by selecting $x$ in
(\ref{special x}). However, this choice is inconsistent with a zero
beta function for the cosmological constant. Let us check this
statement. The new modified beta functions will have the following form:
\beq
\tilde{\be}_C &=&
 \frac{1}{(4\pi)^{2}} \left( \frac{2x}{9}+\frac{397}{40} + c_C \right),
\label{betaW2}
\\
\tilde{\be}_{R} &=&
 \frac{1}{(4\pi)^{2}} \left( -\frac{7}{36} + c_R \right) ,
\label{betaR22}
\\
\tilde{\be}_{\rm GB} &=&
\frac{1}{(4\pi)^{2}} \,\left( \frac{1387}{180} + c_{\rm GB} \right),
\label{betaGB2}
\\
\tilde{\be}_{\kappa} &=&
  \frac{1}{(4\pi)^{2}} \,
\left[ -\frac{5 \theta_C(t)}{6 \omega_C} - \frac{\theta_R(t)}{2 \omega_R}
+  \frac{5 \theta_R(t)}{ \omega_C}+c_\ka\right]
\label{betaR2-a} ,
\\
\be_{\Lambda} &=&
\frac{1}{(4\pi)^{2}} \,
\left[
\frac{5 \omega_\kappa(t) }{2 \omega_C}
- \frac{\omega_\kappa(t) }{6 \omega_R}
- \frac{5}{2} \left( \frac{\theta_C(t)}{\omega_C}  \right)^2
- \frac{1}{2} \left( \frac{\theta_R(t)}{\omega_R}  \right)^2
\right] .
\label{betaLambda-a}
\eeq
Here $c_i=(c_C,\, c_R,\,c_{\rm GB},\,c_\ka)$ are contributions coming from the
cubic in curvature nonminimal terms. All such $c_i$ can be, in principle, made arbitrary by choosing the corresponding coefficients in the classical action.

It is clear that proper choices of $c_C$, $c_R$
can make zero the beta functions $\beta_C$ and $\beta_R$. We analyze such a situation here.
Therefore, we can consistently take the initial conditions $\theta_C(0)$ and $\theta_R(0)$  to be
zero because, in this case, the theory is finite for what concerns the operators $C^2$ and $R^2$ in the
divergent action (\ref{Ga}). This also means that these two terms can be removed completely from the classical action (\ref{L2})
and consistently they will not be recreated by quantum corrections ($\th_C(t)=\th_R(t)=0$).
Moreover, we can replace
\beq
\omega_C = 15 \omega_R, \quad \implies \quad x=15,
\label{Lzero}
\eeq
in (\ref{betaLambda-a}) to finally get $\beta_\Lambda = 0$. Hence,
we can fix $\Lambda =0$ consistently at any energy scale, at the one-loop level.
Under these conditions, at the end, only $\omega_\kappa(t)>0$ can run and there we have asymptotic
freedom as elucidated above. By adjusting to zero the value of the $c_\ka$ coefficient, one can make
absent also the running of $\omega_\kappa$ and the theory is completely UV-finite.
Finally, the relations (\ref{C1L}) and (\ref{C2L}) turn into
\beq
&& \Delta_2 = - 18 \omega_C \,  \omega_\kappa<0  \, , \label{condition1}
\\
&& \Delta_0 =  6 \omega_R \, \omega_\kappa <0 \, . \label{condition2}
\eeq
From (\ref{Lzero}), it is obvious that we must consider only two cases: either $\om_C,\om_R>0$ or
$\om_C,\om_R<0$.
In the first case,
the spectrum consists of the graviton, two LW-like spin-two
fields (from $\Delta_2<0$) with purely imaginary mass square parameter, one real scalar
and a scalar tachyon (from $\Delta_0>0$). Thus, the theory is unitary according to
Anselmi-Piva's prescription, but it is unstable because of the presence of the
tachyon.
On the other hand, if we take $\omega_C,\omega_R < 0$ the situation is reversed. We end up with
 the poles of the ghosts in the spin-two sector becoming real: one ghost field and a ghost tachyon.
 In the spin-zero sector we meet  two LW-like
fields with purely imaginary mass square parameter. One concludes that the conditions (\ref{condition1}) and (\ref{condition2}),
together with (\ref{Lzero}) are inconsistent, mainly due to the presence of ghosts and tachyons.
When one allows for the running of the Newton constant $\om_\ka\to\om_\ka(t)$,
the situation does not get any better.

Therefore, in order to avoid tachyons, one may try to make
the cosmological constant nonzero and make it to run with the energy scale as well as the Newton constant,
according to the following formulas:
\beq
&&
\omega_\La(t) = \omega_\Lambda(0)
+ \frac{1}{ (4 \pi)^2}
\left( \frac{5}{2 \omega_C} -\frac{1}{6 \omega_R} \right) \left(\om_\ka(0)\,t+\frac{c_\ka}{2(4\pi)^2}\,t^2\right),\label{omLarun}
\\
&&
\omega_\kappa(t) = \omega_\kappa(0)
+ \frac{1}{ (4 \pi)^2} c_\kappa \, t \,. \label{omkrun}
\eeq
We still assume that $\th_C(t)=\th_R(t)=0$ both in the classical action and in its RG-improved quantum version. The ghost poles are now located in the following points (in accordance with (\ref{H2z}) and (\ref{H0z})):
\beq
&&
\mbox{spin-2}: \quad  z_{1, 2}^{(2)}
= \frac{ \pm \sqrt{\Delta_2}}{6 \omega_C} ,
\\
&&
\mbox{spin-0}: \quad z_{1,2}^{(0)}
=  \frac{ 4 \Lambda \omega_R \pm  \sqrt{ \Delta_0 }}{6 \omega_R} .
\eeq
In order to avoid tachyons in the spin-0 sector case we must take $\Lambda<0$
 to have negative real contribution to the mass squared, while
the spin-2 sector is without this problem. The two fields of spin-2 have both zero real parts of
the mass square parameter, which is then a purely imaginary quantity.
We assume that $\om_C>0$, $\om_R<0$ (or $\om_R>1/15\,\om_C$) and also that $\om_\La(0)>0$. Moreover, $\om_\ka(0)>0$  and $c_\ka>0$, so $\om_\ka(t)>0$.
Note that now the running parameter $\om_\Lambda(t)>0$ and hence the cosmological constant $\La(t)$ is negative according to (\ref{cosmoomega}), while the
constraints (\ref{C1L}) and (\ref{C2L}) simplify to
 \beq
&&
\Delta_2 =
( 8 \Lambda(t) \omega_C)^2 - 18 \omega_C \,\omega_\kappa(t)  < 0,
\label{C1L3}
\\
&&
\Delta_0 =
2   \omega_R  \left( 3 \omega_\kappa(t) +8\La(t)^2\om_R \right)< 0 .
\label{C2L3}
\eeq
Both constraints cannot
be satisfied since both $\Delta_2<0$ in (\ref{C1L3}) and $\Delta_0<0$ in (\ref{C2L3})
effectively lead in the large $t$ leading asymptotics to the expression
\beq
\om_C^2 - 30\, \om_C \om_R + 225\, \om_R^2 = \left( \om_C - 15\, \om_R\right)^2 \geqslant0,
\eeq
which obviously cannot be negative for any $\om_C$ and $\om_R$. This is
a simple consequence of the fact that in (\ref{C1L3}) and  in (\ref{C2L3}) the
$\La(t)^2$ term has the leading $t^2$ asymptotics  higher than that of $\om_\ka(t)\sim t$ (cf. (\ref{omkrun})).
One can also consider the situation where only one of the couplings $\om_\La$ or $\om_\ka$
shows quantum renormalization running. In the first case, when $c_\ka=0$,
the running $\om_\La(t)$, which scales like $t$ according to (\ref{omLarun}),
again dominates the relations (\ref{C1L3}) and (\ref{C2L3}) in the UV regime
of large $t$ due to the fact that $\La(t)\sim t$. Now repeating verbatim the same
argumentation as above, one sees that it is impossible to satisfy
both $\Delta_2<0$ in (\ref{C1L3}) and $\Delta_0<0$ in (\ref{C2L3}) for all
values of $t$. When $\om_C=15\om_R$ and $c_\ka\neq0$, the coupling $\om_\ka$ is the
only one which runs and it is linear in $t$. In this situation, $\om_\ka(t)$ is the
dominant quantity for large $t$ in relations  (\ref{C1L3}) and  (\ref{C2L3}), so
effectively we can forget about $\Lambda(t)$ there (which falls off here like $1/t$).
This means that in the large $t$ asymptotics we are back to relations
(\ref{condition1}) and  (\ref{condition2}) obtained for
$\La(t)=0$ that we showed above cannot be satisfied simultaneously
together with $\om_C=15\om_R$.

In the final comment, we can return to the case when the full theory
is completely UV-finite, but when we keep a nonzero value of the $\om_\La$
coupling. The choice of signs explained above implies that $\La<0$.
A careful analysis of the inequalities (\ref{C1L3}) and  (\ref{C2L3})
with all the couplings taking constant values and satisfying the
relation $\om_C=15\om_R$ shows that again it is impossible to
find a nonzero allowed interval for values of the $\om_R$ coupling. Simply, restricting
to $\om_R>0$, the $\Delta_2<0$ implies $\om_R<\frac{3}{160} \om_\ka \La^{-2}$, while
on the other side
$\Delta_0<0$ implies  that $\om_R>\frac{3}{8} \om_\ka \La^{-2}$.
One therefore concludes that some of the couplings $\th_C$ or $\th_R$
must run with energy too to provide enough room for keeping the pair
of Lee-Wick particles with complex conjugate poles at any energy scale.

Eventually, the full theory consistent with the analysis in the last paragraph
will include all or some of the following cubic curvature invariants, written in
general dimensions
\beq
&&\hspace{-1cm}R^{3},\quad RR_{\mu\nu}R^{\mu\nu},\quad R_{\mu\nu}R^{\mu}{}_{\rho}R^{\nu\!\rho},\\
&&\hspace{-1cm}RR_{\mu\nu\!\rho\sigma}R^{\mu\nu\!\rho\sigma},\quad R_{\mu\nu}R_{\rho\sigma}R^{\mu\!\rho\nu\sigma},\quad
R_{\mu\nu}R^{\mu}{}_{\rho\si\ka}R^{\nu\!\rho\si\ka},\quad R_{\mu\nu\!\rho\sigma}R^{\mu\nu}{}_{\kappa\lambda}R^{\rho\sigma\kappa\lambda},\quad
R_{\mu\!\rho\sigma\nu}R^{\rho\kappa\lambda\sigma}R_{\kappa}{}^{\mu\nu}{}_{\lambda}
\,,\label{Riem2}
\eeq
but due to various identities holding in four spacetime dimensions \cite{MPR2020}
only three out of the five operators containing a Riemann tensor in (\ref{Riem2}) are
independent there. Therefore we can add the following six nonminimal terms below, which are cubic in gravitational curvatures with arbitrary coefficients $s_1,\ldots,s_6$.  The explicit form of the addition to the Lagrangian in (\ref{L}) reads\footnote{One notices that the generalized Gauss-Bonnet scalar ${\rm GB}_1$ (\ref{GB1}) is not independent of the terms written above since it gives rise to precise values of the coefficients $s_1,\ldots,s_6$ in four dimensions after exploiting integration by parts under spacetime volume integral, commutation of covariant derivatives and Bianchi and cyclic identities for the Riemann tensor.}
\beq
 s_1R^{3}+s_{2}RR_{\mu\nu}R^{\mu\nu}+s_{3}R_{\mu\nu}R^{\mu}{}_{\rho}R^{\nu\!\rho}
 +s_4RR_{\mu\nu\!\rho\sigma}R^{\mu\nu\!\rho\sigma} +s_{5}R_{\mu\nu}R_{\rho\sigma}R^{\mu\!\rho\nu\sigma}+s_6R_{\mu\nu\!\rho\sigma}R^{\mu\nu}{}_{\kappa\lambda}R^{\rho\sigma\kappa\lambda}\,.
\eeq
 With these additional operators at our disposal, we
will likely get the modified beta functions in the general form (\ref{betaW2})-(\ref{betaR2-a}).

One can notice the following schematic relations of how the running couplings depend on the constant parameters $s_i$ with $i=1,\ldots,6$ (where we keep only the highest powers). First, the couplings $\th_C(t)$,  $\th_R(t)$ and $\th_{\rm GB}(t)$ show up to quadratic dependence on $s_i$ (so they are proportional to $s_i s_j$ in the numerators). The denominators of such ratios can be of three different types: $\omega_C^2$, $\omega_R^2$ or $\omega_C \omega_R$. The running Newton coupling $\omega_\kappa(t)$ will generally depend on a cubic form in $s_i$ coefficients in the numerators, while the running cosmological constant $\omega_\Lambda(t)$ on up to a quartic one.

One last comment concerns our taken for granted trust in the energy dependence of the spectrum. Consistently with the perturbative expansion, the classical spectrum cannot be largely modified in the loop expansion. However, asymptotic freedom allows us to infer about the ultraviolet regime of the theory (mathematically this is the infinite energy regime). Therefore, we are forced to consider the spectrum at any energy scale with particular attention to the UV regime.

\section{Conclusions and discussions}
\label{s7}

For the first time, we derived the exact beta functions for the
four-derivative coefficients in the minimal, super-renormalizable
model of the six-derivative QG. This minimal theory is the simplest
super-renormalizable model, with only seven free parameters in the
classical action (\ref{L}).

The result is applicable to the minimal Lee-Wick
quantum gravity proposed in \cite{LM-Sh,ModestoLeeWick},
which has been shown to be unitary at any perturbative order
and also admits the  procedure suggested in
\cite{AnselmiPiva1, AnselmiPiva2, AnselmiPiva3}.
For this model, we solved the renormalization group equations,
showing asymptotic freedom and two possible perturbative
scenarios in the ultraviolet regime:
in one case the cosmological
constant runs towards zero, while in the other case it has a
negative-value fixed point.
In the former scenario a free graviton propagates in Minkowski spacetime, while in the latter one the graviton propagates in the (anti-) de Sitter spacetime, which may be unstable in the minimal six-derivative model because of the presence of tachyons
in the spectrum.
 Therefore, we
proposed an extension of the theory that very likely will be free
of tachyons, while keeping asymptotic freedom on the (A)dS
space in the ultraviolet regime.

Our novel results for UV divergences (proportional to $C^2$, $R^2$ and the $E_4$ term) obtained
in this work are completely free of any problem of gauge dependence. We emphasize that these results are
gauge-independent and also gauge-fixing-independent. This is another remarkable and very positive and
physically welcomed feature of our model. And it is in opposition for example to the case of the beta
function of the $G_N$ coupling in Stelle's quadratic theory in $d=4$ spacetime dimensions, which still shows
various ambiguities  \cite{SRQG-beta}. The UV divergences that we have found are fully unambiguous,
gauge-independent and they represent true physical quantities since they have the meaning not only in the
perturbative approach as one-loop expressions, but also they are perturbatively exact (as also explained
below). The mathematical fact that these UV divergences are gauge-independent is proven by the powerful
theorem (originally due to Kallosh, Tarasov, and Tyutin in \cite{KTT}) that we have already explained in
the point iii) of Sec.~\ref{s2}. Hence there is not any problem with gauge dependence of our results.

Similarly, we just remark for completeness that the locations of the poles in both spin-2 and spin-0
parts of the gravitational propagator of the theory both around flat and (A)dS backgrounds are completely
independent of the gauge choices used to compute the propagator. Hence also the masses of spin-0 and
spin-2 particles present in the spectrum are completely gauge-fixing independent. This fact was crucial
for the consistency and the physical sense of the analysis  that we have presented in Sec.~\ref{s5} and in Sec.~\ref{s6}. This
is in accordance with the existing literature on the topic.

We also put special attention to the fact that the condition for AF depends crucially on the initial
conditions of the RG flow and also on the signs of the beta functions. This point we explained at length
also before. In many places above in Sec.~\ref{s6}, we gave reasonable physical inequalities for the signs of the initial
values of the couplings. For example, we required that $\omega_R(0)<0$, $\omega_C(0)>0$,
$\omega_{\rm GB}(0)>0$, etc. We also demanded for consistency and AF in the UV the corresponding
 matching signs of the beta functions such that the RG flow for all finite values of $t$ is regular
 and does not touch neither zero nor infinite values.

We would like to remind the reader that exactness of our beta functions has to do with the absence of
perturbative loop divergences at loop levels higher than the first one. Simply the theory is
super-renormalizable. This conclusion is based on power counting analysis of UV divergences.

In this gravitational model with six derivatives, one is sure that besides the one-loop level there
are no new perturbative UV divergences proportional to the $C^2$, $R^2$ and the $E_4$ terms.
Since one reads the beta functions of the couplings of the theory
from these divergences, then one concludes that expressions for the beta functions at the one-loop level
take into account all loop contributions and hence corresponds to full loop resummation. Assuming that
there are no nonperturbative contributions to the beta functions of couplings, then this means that the
beta functions that we have obtained are exact and there will be no additional contribution to them
due to any other quantum effects. Just one-loop quantum effects encoded in UV divergences are completely
enough to settle the exact form of the RG running of couplings  $\theta_C$, $\theta_R$ and $\theta_{\rm GB}$ in this model.

Moreover, the structure of
the beta functions for $\omega_\kappa$ and for $\omega_\Lambda$ as written in (\ref{betaR}) and
(\ref{betaLambda}) respectively must be repeated also when two-, two- and three-loop contributions are
taken additionally into account there. This is due to an expected form of divergences at two and three loop
orders, which compared to formulas in (\ref{betaR}) and (\ref{betaLambda}) from dimensional reasons can
be only multiplied by the powers (both positive and negative) of the fundamental ratio $x$ of the theory.
Hence the integration of these formulas leads to the running the same as encoded in formulas
(\ref{runningkappa}) and (\ref{runningLambda}) respectively, where only the constant numerical coefficients may be
different after inclusion of higher loop effects. But the leading $t^2$ and $t^3$ behaviors respectively  for the
running coupling parameters
$\omega_\kappa(t)$ and for $\omega_\Lambda(t)$ in the UV regime are a universal feature here. Together with exactness of one-loop expressions for running $\theta_C(t)$,
$\theta_R(t)$ and $\theta_{\rm GB}(t)$, this is one of the most beautiful and powerful features of the
super-renormalizable model we have considered in this work. Therefore this model constitutes a good and promising theoretical laboratory for studying RG flows.

Let us mention two possible extensions of the results described in
this paper.

The beta functions (\ref{betaW}), (\ref{betaR2}), (\ref{betaGB}),
as well as the beta functions for the linear in curvature term and
the complete one-loop beta function for the cosmological constant, can be used
for testing alternative approaches to QG. For instance, there are
interesting approaches, that are regarded nonperturbative, such as
functional renormalization group or dynamical triangulations. It
would be interesting to see whether these methods can reproduce
the beta functions that were derived here. Such a comparison
represents one of the possibilities to extend the present work.

Moreover, this paper lays the foundations for future computations
with more involved (one can say nonminimal) models for a
super-renormalizable QG. As we have discussed above, there are
many possible extensions of the minimal Lagrangian (\r{L}). In
this respect, the situation is very similar to the minimal Standard Model
of particle physics. However, the important difference is that for
the QG models there are no experimental or phenomenological
restrictions on the higher derivative parameters of the simplest
model (\r{L}), or on its nonminimal extensions. On the other hand,
it would be interesting to explore the IR limit of the theory with
complex poles without or with loop corrections, e.g., in the way it
was done recently in \cite{ABSh1}. In principle, a  detailed analysis
of this aspect of the models under discussion could be useful for
many phenomenological applications.

\section*{Appendix A: Wick rotation in a general local higher derivative theory}
\label{app}
Here we address the issue of Wick rotation in higher derivative theory.
The RG equations are by definition obtained in the deep UV regime of the Euclidean theory. And the question how to rotate them back to Minkowskian/Lorentzian signature has been addressed in Appendix A of Ref. \cite{Aglietti:2016pwz}. Let us here expand the latter proof.

First of all let us remind the reader of the main feature of dimensional regularization. We can consider the following prototype of a one-loop integral suitable for a general higher derivative theory,
\beq
&& \int \frac{d^D k }{(2 \pi)^D} \frac{(k^2)^r}{(k^2 + m^2)^s}
= i \frac{ (m^2)^{ \frac{D}{2} + r - s}}{ (4 \pi)^{\frac{D}{2}}}
\frac{ \Gamma(s - \frac{D}{2} - r) \, \Gamma(\frac{D}{2} + r)}{\Gamma(\frac{D}{2}) \Gamma(s)} \,  ,
\label{one-Loop-integrals}
\eeq
where $s$ and $r$ are integers. We can consider the case in which $r=0$ and $s = - N$ and define the following integral in dimensional regularization:
\beq
\int \frac{d^D k }{(2 \pi)^D} k^{2 N} \equiv
\lim_{m\rightarrow 0} \int \frac{d^D k }{(2 \pi)^D} (k^2 + m^2)^N
\, .
\label{mto0}
\eeq
In order to compute the above integral we replace $r=0$ and $s = - N$ on the right side of (\ref{one-Loop-integrals}), namely
\beq
&& \int \frac{d^D k }{(2 \pi)^D} \frac{1}{(k^2 + m^2)^{-N}} =
 i \frac{ (m^2)^{ \frac{D}{2} + N }}{ (4 \pi)^{\frac{D}{2}}}
\frac{ \Gamma(-N - \frac{D}{2}) \, \Gamma(\frac{D}{2})}{\Gamma(\frac{D}{2}) \Gamma(-N)} \,  .
\label{one-Loop-inetgrals2}
\eeq
For $N+D/2>0$ (strictly) or  $N> - D/2$ the Gamma functions in (\ref{one-Loop-inetgrals2}) are singular, but the limit in (\ref{mto0}) is well defined and it gives zero for $N>-D/2$, i.e.
\beq
\lim_{m\rightarrow 0} \int \frac{d^D k }{(2 \pi)^D} (k^2 + m^2)^N = 0 \quad {\rm for}\quad N>- \frac{D}{2} \, .
\label{mto02}
\eeq
Other more rigorous proofs can be found in \cite{DimRegGen,AnselmiBook}. In short, following  \cite{AnselmiBook}, the integrals in (\ref{one-Loop-inetgrals2}) can be defined as sums of other integrals convergent in a proper domain of the complex variable $D$ (the dimension of spacetime). The result of these integrals turns out to be analytic in $D$ and, therefore, its uniqueness allows us to compare integrals initially defined in different domains. The values of such integrals can finally be combined to reconstruct the integrals (\ref{one-Loop-inetgrals2}), which turn out to be zero.

We now apply the result (\ref{mto02}) to the integrals on the arcs in the first and third quarter of the energy complex plane and we show that they vanish in dimensional regularization. Once again, this is a general result true in any regularization scheme, but easy  to prove in dimensional regularization.

Following the discussion in \cite{Aglietti:2016pwz}, the general one-loop integral (\ref{one-Loop-integrals}) can be written in the following form:
\beq
\hspace{-0.4cm}\int \frac{d^D k }{(2 \pi)^D} \frac{(k^2)^r}{(k^2 + C)^s}=
 \int_{- \infty}^{+ \infty} \frac{d E}{2 \pi} \int \frac{d^{D-1} {\bf k} }{(2 \pi)^{D-1}}
\, \frac{(k^2)^r}{ ( k^{2} + C^2)^s} =
\int_{- \infty}^{+ \infty} \frac{d E}{2 \pi} \int \frac{d^{D-1} {\bf k}}{(2 \pi)^{D-1}} \,  \frac{E^{2 r} +  \sum_{j=1}^{2 r} a_j( {\bf k}) \, E^{2 r - j}  }{E^{2 s} + \sum_{i=1}^{2 s} b_i(p, {\bf k}) \, E^{2 s -i}}  ,
\label{ekint}
\eeq
where here $C$ is a scalar function of the external energy $p_0$ and momenta $\bf p$ combined in a Lorentz-covariant vector $p=(p_0, \bf{p})$ and possibly of  masses, $a_j({\bf k})$ is a polynomial of ${\bf k}$
and $b_i(p, {\bf k})$ is polynomial of ${\bf k}$ and the external $D$-dimensional momenta $p$, while $s\geqslant1$ and $r\geqslant0$ are integers. (Similarly, we have the $D$-dimensional Lorentz vector $k=(E,\bf{k})$ over which we integrate.)
The energy integral in (\ref{ekint}) is divergent for $2 r + 1\geqslant 2 s$, hence, we define $2 r = 2 s + n -1$ and take $n\geqslant 0$.
In order to compute the contribution to the integral (\ref{ekint}) along the arcs, let us take $|E|$ much larger than all the other scales in the integrand. Therefore, the divergent contributions are obtained expanding the denominator in (\ref{ekint}) as follows:
\beq
&&  \int_{| E | \sim \infty} \frac{d E}{2 \pi} \int \frac{d^{D-1} {\bf k}}{(2 \pi)^{D-1}} \,
E^{n-1} \left[
1 - \sum_{i=1}^{2 s} \sum_{j=1}^{2 s + n -1} \frac{ b_i(p, {\bf k}) a_j( {\bf k})}{E^{i+j} }
+
\sum_{i=1}^{2 s} \sum_{k=1}^{2 s} \sum_{j=1}^{2 s + n -1} \frac{ b_i(p, {\bf k}) b_k (p, {\bf k}) a_j( {\bf k})  }{E^{i+k+j} } + \dots
\right] \, .
\label{ekintT}
\eeq
Notice that the integral in the energy $E$ is actually a convergent angular integral multiplied by the radius $|E|$. Therefore, we have to focus only on integrals that diverge in the limit $|E| \rightarrow \infty$ after the angular integration has been performed. Since the theory is local, the number of such divergent integrals is finite and, therefore, we can add a finite number of integrals in $(D - 1)$ dimensions that are ``zero'' as a particular feature of the dimensional regularization in $(D-1)$ spacetime dimensions.
Indeed, all such integrals are polynomial integrals of ${\bf k}$ that turn out to be zero when the dimensional regularization (\ref{mto02}) is implemented in the $(D-1)$-dimensional manifold.
Therefore, whenever we need to close the integration path, in the lower or upper half plane, we actually add a finite sum of vanishing contributions.

Let us expand about this statement.
In general, we add to the integral (\ref{ekint}) a finite number of ``zeros'' by  means of using the dimensional regularization (DIMREG) scheme. However, for the sake of simplicity, we here consider only the case $r=s$ or $n=1$ (for Einstein's gravity $r=s=1$), hence, in
the expansion (\ref{ekintT}) only the first volume contribution is divergent,
\beq
&& \int \frac{d^D k }{(2 \pi)^D} \frac{(k^2)^r}{(k^2 + C)^r}=
 \int_{- \infty}^{+ \infty} \frac{d E}{2 \pi} \int \frac{d^{D-1} {\bf k} }{(2 \pi)^{D-1}}
\, \frac{(k^2)^r}{ ( k^{2} + C^2)^r} =
\int_{- \infty}^{+ \infty} \frac{d E}{2 \pi} \int \frac{d^{D-1} {\bf k}}{(2 \pi)^{D-1}} \,  \frac{E^{2 r} +  \sum_{j=1}^{2 r} a_j( {\bf k}) \, E^{2 r - j}  }{E^{2 r} + \sum_{i=1}^{2 r} b_i(p, {\bf k}) \, E^{2 r -i}}  \nonumber \\
&& =
 \int_{| E | \sim \infty} \frac{d E}{2 \pi} \int \frac{d^{D-1} {\bf k}}{(2 \pi)^{D-1}} \,
 \left[
1 - \sum_{i=1}^{2 s} \sum_{j=1}^{2 s } \frac{ b_i(p, {\bf k}) a_j( {\bf k})}{E^{i+j} }
+
\sum_{i=1}^{2 s} \sum_{k=1}^{2 s} \sum_{j=1}^{2 s} \frac{ b_i(p, {\bf k}) b_k (p, {\bf k}) a_j( {\bf k})  }{E^{i+k+j} } + \dots
\right] \nonumber \\
&&
=
 \int_{| E | \sim \infty} \frac{d E}{2 \pi} \int \frac{d^{D-1} {\bf k}}{(2 \pi)^{D-1}} \,
 \left[ \sum_{\ell=0}^{+ \infty}
\left( -1 \right)^\ell
\left(  \sum_{i=1}^{2 r} \frac{b_i(p, {\bf k})}{E^{i}} \right)^\ell
\right]
\left[ 1 +  \sum_{j=1}^{2 r} \frac{a_j( {\bf k})}{E^{ j}}  \right]
- \underbrace{\int_{| E | \sim \infty} \frac{d E}{2 \pi} \int \frac{d^{D-1} {\bf k}}{(2 \pi)^{D-1}}}_{= 0 \,\, \mbox{in DIMREG}}
\nonumber \\
&&
=
 \int_{| E | \sim \infty} \frac{d E}{2 \pi} \int \frac{d^{D-1} {\bf k}}{(2 \pi)^{D-1}} \,
\left\{ \left[ \sum_{\ell=0}^{+ \infty}
\left( -1 \right)^\ell
\left(  \sum_{i=1}^{2 r} \frac{b_i(p, {\bf k})}{E^{i}} \right)^\ell
\right]
\left[ 1 +  \sum_{j=1}^{2 r} \frac{a_j( {\bf k})}{E^{ j}}  \right]
- \underbrace{1}_{= 0 \,\, \mbox{in DIMREG}} \right\}
\nonumber \\
&&
=
 \int_{| E | \sim \infty} \frac{d E}{2 \pi} \int \frac{d^{D-1} {\bf k}}{(2 \pi)^{D-1}} \,
\left\{ \left[ \sum_{\ell=1}^{+ \infty}
\left( -1 \right)^\ell
\left(  \sum_{i=1}^{2 r} \frac{b_i(p, {\bf k})}{E^{i}} \right)^\ell
\right]
\left[ 1 +  \sum_{j=1}^{2 r} \frac{a_j( {\bf k})}{E^{ j}}  \right]
 \right\}
= 0 \quad {\mbox{for}} \quad |E| \rightarrow + \infty
\, .
\label{ekint2}
\eeq
Now, we observe that when we take the limit $|E| \rightarrow + \infty$ (large radius in the first and third sector in the energy complex plane)
only the first integral in the sum (\ref{ekint2}) diverges.
However, it is identically zero or, which is equivalent, it cancels with the introduced ``zero'', as a particular feature of dimensional regularization. Notice that in the last step we did not use any expansion, but only the limit for large $|E|$ in the first of the two integrals. Of course, the sum of two integrals is the sum of the integrands. \\
In the general case $2 r + 1\geqslant 2 s$ and introducing the definition $2 r = 2 s + n -1$, the integral reads:
\beq
&& \int \frac{d^D k }{(2 \pi)^D} \frac{(k^2)^r}{(k^2 + C)^s}=
 \int_{- \infty}^{+ \infty} \frac{d E}{2 \pi} \int \frac{d^{D-1} {\bf k} }{(2 \pi)^{D-1}}
\, \frac{(k^2)^r}{ ( k^{2} + C^2)^s} =
\nonumber \\
&& =
\int_{- \infty}^{+ \infty} \frac{d E}{2 \pi} \int \frac{d^{D-1} {\bf k}}{(2 \pi)^{D-1}} \,  \frac{E^{2 r} +  \sum_{j=1}^{2 r} a_j( {\bf k}) \, E^{2 r - j}  }{E^{2 s} + \sum_{i=1}^{2 s} b_i(p, {\bf k}) \, E^{2 s -i}}  \nonumber \\
&& =
\int_{- \infty}^{+ \infty} \frac{d E}{2 \pi} \int \frac{d^{D-1} {\bf k}}{(2 \pi)^{D-1}} \, E^{2 r-2 s} \,
\left[ \frac{1 +  \sum_{j=1}^{2 r} a_j( {\bf k}) \, E^{ - j}  }{1 + \sum_{i=1}^{2 s} b_i(p, {\bf k}) \, E^{ -i}} \right]
\nonumber \\
&&
\equiv
\int_{- \infty}^{+ \infty} \frac{d E}{2 \pi} \int \frac{d^{D-1} {\bf k}}{(2 \pi)^{D-1}} \, E^{n-1} \,  \left[ \frac{1 +  \sum_{j=1}^{2 r} a_j( {\bf k}) \, E^{ - j}  }{1 + \sum_{i=1}^{2 s} b_i(p, {\bf k}) \, E^{ -i}} \right]
\nonumber \\
&&
= \int_{| E | \sim \infty}  \frac{d E}{2 \pi} \int \frac{d^{D-1} {\bf k}}{(2 \pi)^{D-1}} \, E^{n-1} \,  \left[ \frac{1 +  \sum_{j=1}^{2 r} a_j( {\bf k}) \, E^{ - j}  }{1 + \sum_{i=1}^{2 s} b_i(p, {\bf k}) \, E^{ -i}} \right] \nonumber\\
&&
=
 \int_{| E | \sim \infty} \frac{d E}{2 \pi} \int \frac{d^{D-1} {\bf k}}{(2 \pi)^{D-1}} \,
E^{n-1} \,  \left[ \sum_{\ell=0}^{+ \infty}
\left( -1 \right)^\ell
\left(  \sum_{i=1}^{2 r} \frac{b_i(p, {\bf k})}{E^{i}} \right)^\ell
\right]
\left[ 1 +  \sum_{j=1}^{2 r} \frac{a_j( {\bf k})}{E^{ j}}  \right]
\nonumber \\
&&
=
 \int_{| E | \sim \infty} \frac{d E}{2 \pi} \int \frac{d^{D-1} {\bf k}}{(2 \pi)^{D-1}} \,
E^{n-1} \left[
1 - \sum_{i=1}^{2 s} \sum_{j=1}^{2 s + n -1} \frac{ b_i(p, {\bf k}) a_j( {\bf k})}{E^{i+j} }
+
\sum_{i=1}^{2 s} \sum_{k=1}^{2 s} \sum_{j=1}^{2 s + n -1} \frac{ b_i(p, {\bf k}) b_k (p, {\bf k}) a_j( {\bf k})  }{E^{i+k+j} } + \dots
\right]\nonumber
\\
&& +
 \int_{| E | \sim \infty} \frac{d E}{2 \pi} \,  E^{n-1} \underbrace{\int \frac{d^{D-1} {\bf k}}{(2 \pi)^{D-1}} \, 1}_{= 0 \,\, \mbox{in DIMREG}}
 + \, c_1 \, p^2
 \int_{| E | \sim \infty} \frac{d E}{2 \pi} \,  E^{n-3} \underbrace{\int \frac{d^{D-1} {\bf k}}{(2 \pi)^{D-1}} \, 1}_{= 0 \,\, \mbox{in DIMREG}}
+ \, c_2
 \int_{| E | \sim \infty} \frac{d E}{2 \pi} \,  E^{n-3} \underbrace{\int \frac{d^{D-1} {\bf k}}{(2 \pi)^{D-1}} \, {\bf k}^2}_{= 0 \,\, \mbox{in DIMREG}}
 \nonumber \\
&& + \ldots +
 \, c_n
 \int_{| E | \sim \infty} \frac{d E}{2 \pi} \,  E^{-1} \underbrace{\int \frac{d^{D-1} {\bf k}}{(2 \pi)^{D-1}} \, {\bf k}^{2 n}}_{= 0 \,\, \mbox{in DIMREG}}
 +  \, O\left( \frac{1}{E} \right)
 = 0 \quad {\mbox{for}} \quad |E| \rightarrow + \infty
\, ,
\label{ekint3}
\eeq
where $c_1, \dots, c_n$ are constants.

Coming back to our calculation in the main text of the paper, it was done exclusively and from the beginning
in the Euclidean domain (because of using Euclidean-signature based  Barvinsky-Vilkovisky trace technology).

Afterwards, in order to get the theory in Minkowski spacetime, we need the Anselmi-Piva (AP) prescription \cite{AnselmiPiva1,AnselmiPiva2,AnselmiPiva3} for which it is technically crucial that the contributions on the arcs discussed in this appendix vanish in DIMREG in $(D-1)$ dimensions.
However, such prescription does not affect the divergences that are computed in Euclidean signature.
Using the AP prescription the divergences are the same in Euclidean as
in the Lorentzian case without any need to do explicitly Wick rotation.
One could even say that the UV divergences that we have obtained as considered
 in Lorentzian theory are independent of any prescription
scheme of how to encircle/include/exclude poles of the propagator in the Euclidean theory
and this issue of how to close the integration contour on the complex plane is not essential
for performing formal Wick rotation to the Minkowskian domain.
This is so since these are subleading effects (they do matter for finite terms,
but not for the leading in the UV terms of the expansion of loop integrals).
Hence the UV divergences in the form we have found are the same in Euclidean
as well as in Lorentzian signature. Here we assume that there exists a well defined
calculational procedure in the Euclidean regime, or that the same calculations can be equivalently performed using Feynman diagrams,
but for the UV divergences the results will be the same. The situation is a verbatim repetition of
the case of beta functions of higher derivative Stelle's quadratic theory \cite{Stelle} in $D=4$ spacetime dimensions.
Also there the beta functions are valid universally both in Euclidean and Minkowskian
theory. Of course, in the latter case, one has to understand the various curvature tensors (Riemann, Weyl, etc.) as computed using the spacetime metric with Lorentzian signature as opposed to the former case of Euclidean theory, where they are computed on the metric with completely positive Euclidean signature.


\end{document}